\documentclass[twocolumn]{aastex63}
\usepackage{amsmath}
\usepackage{soul}
\usepackage{lineno}

\newcommand\Mstar{\rm M_{\star}}
\newcommand\Rstar{\rm R_{\star}}
\newcommand\Msun{\rm M_{\odot}}

\newcommand\Prot{P_\mathrm{{rot}}}
\newcommand\Psat{P_\mathrm{{sat}}}
\newcommand\tauc{\tau_\mathrm{c}}
\newcommand\HP{H_\mathrm{P}}
\newcommand\vc{v_\mathrm{c}}
\newcommand\taucE{\tau_\mathrm{cE}}
\newcommand\taucM{\tau_\mathrm{cM}}
\newcommand\Ro{\mathrm{Ro}}
\newcommand\Lx{L_\mathrm{X}}
\newcommand\Lbol{L_\mathrm{bol}}
\newcommand\rbce{r_\mathrm{BCE}}
\newcommand\bruntvaisala{Brunt-V\"{a}is\"{a}l\"{a}}

\submitjournal{ApJ}

\shorttitle{Convective Turnover Times in Low-Mass Stars}
\shortauthors{Gossage et al.}
\graphicspath{{./}{figures/}}
\begin{document}
%\linenumbers

\title{On Convective Turnover Times and Dynamos In Low-Mass Stars}

\correspondingauthor{Seth Gossage}
\email{seth.gossage@northwestern.edu}

\author[0000-0001-6692-6410]{Seth Gossage}
\affiliation{Center for Interdisciplinary Exploration and Research in Astrophysics (CIERA), Northwestern University, 2145 Sheridan Road, Evanston, IL 60208, USA}
\affiliation{Center for Astrophysics | Harvard \& Smithsonian, 60 Garden Street, Cambridge, MA~02138, USA}

\author[0000-0003-2102-3159]{Rocio Kiman}
\affiliation{Department of Astronomy, California Institute of Technology, Pasadena, CA 91125, USA}

\author[0000-0002-5688-6790]{Kristina Monsch}
\affiliation{Center for Astrophysics | Harvard \& Smithsonian, 60 Garden Street, Cambridge, MA~02138, USA}

\author[0000-0001-8726-3134]{Amber A. Medina}
\affiliation{Department of Astronomy, The University of Texas at Austin, Austin, TX 78712, USA}
\affiliation{Center for Astrophysics | Harvard \& Smithsonian, 60 Garden Street, Cambridge, MA~02138, USA}

\author[0000-0002-0210-2276]{Jeremy J. Drake}
\affiliation{Lockheed Martin Solar and Astrophysics Laboratory, 3251 Hanover St, Palo Alto, CA 94304}
\affiliation{Center for Astrophysics | Harvard \& Smithsonian, 60 Garden Street, Cambridge, MA~02138, USA}

\author[0000-0002-8791-6286]{Cecilia Garraffo}
\affiliation{Center for Astrophysics | Harvard \& Smithsonian, 60 Garden Street, Cambridge, MA~02138, USA}
\affiliation{Institute for Applied Computational Science, Harvard University, 33 Oxford St., Cambridge, MA 02138, USA}

\author[0000-0003-4769-3273]{Yuxi(Lucy) Lu}
\affiliation{American Museum of Natural History, Central Park West, Manhattan, NY~10024, USA}
\affiliation{Department of Astronomy, The Ohio State University, Columbus, 140 W 18th Ave, OH 43210, USA}

\author[0009-0003-5349-6994]{Joshua D. Wing}
\affiliation{Center for Astrophysics | Harvard \& Smithsonian, 60 Garden Street, Cambridge, MA~02138, USA}

\author[0000-0002-8389-8711]{Nicholas J. Wright}
\affiliation{Astrophysics Group, Keele University, Keele ST5 5BG, UK}

\begin{abstract}
The relationship between magnetic activity and Rossby number is one way through which stellar dynamos can be understood. Using measured rotation rates and X-ray to bolometric luminosity ratios of an ensemble of stars, we derive empirical convective turnover times based on recent observations and re-evaluate the X-ray activity-Rossby number relationship. In doing so, we find a sharp rise in the convective turnover time for stars in the mass range of $0.35{-}0.4\,\Msun$, associated with the onset of a fully convective internal stellar structure. Using \texttt{MESA} stellar evolution models, we infer the location of dynamo action implied by the empirical convective turnover time. The empirical convective turnover time is found to be indicative of dynamo action deep within the convective envelope in stars with masses $0.1{-}1.2\,\Msun$, crossing the fully convective boundary. Our results corroborate past works suggesting that partially and fully convective stars follow the same activity-Rossby relation, possibly owing to similar dynamo mechanisms. Our stellar models also give insight into the dynamo mechanism. We find that empirically determined convective turnover times correlate with properties of the deep stellar interior. These findings are in agreement with global dynamo models that see a reservoir of magnetic flux accumulate deep in the convection zone before buoyantly rising to the surface.

\end{abstract}


\section{Introduction} 
\label{s:intro}

The surge of interest in the nature and conditions of exoplanets and the effects their parent stars have on them has led to a renaissance in the study of the non-thermal emission from stars known very generally as ``stellar activity''. This emission originates in the chromosphere and corona and is nonthermal in the sense that it cannot be explained in terms of the blackbody-like thermal spectra that characterize the photospheric emission of stars in general.

The nature of stellar activity was firmly established as a magnetic phenomenon powered by rotation and convection via an interior dynamo in a series of both solar and stellar studies through the 1970's and early 1980's. Observations of Ca~II H \& K line core chromospheric emission of the Sun revealed a dependence of emission flux on surface magnetic field strength \citep[e.g.][]{Frazier:72}. In stars, H \& K fluxes were found to decrease linearly with projected rotation velocity \citep{Kraft:67}, and with stellar age, $t$, approximately according to $t^{-1/2}$ 
\citep{Skumanich:72} due to gradual angular momentum loss through magnetized stellar winds \citep{Kraft:67,Weber.Davis:67,Durney:72,Mestel.Spruit:87}.
Higher up in the atmosphere, stellar surveys with the {\it Einstein} observatory found that the X-ray luminosity of coronal emission was also highly correlated with stellar rotation period \citep{Vaiana.etal:81,Pallavicini.etal:81,Walter.Bowyer:81}.  

The connection of chromospheric and coronal emission with an interior magnetic dynamo indicates that some fraction of the magnetic energy generated finds its way to the stellar surface, and is dissipated by the observed radiative losses and a presumed stellar wind. At present, none of the processes involved in this chain, from the dynamo itself to chromospheric and coronal heating, are fully understood. The dynamo problem alone encompasses the complicated fluid dynamics of rotating, convective, magnetized plasmas at high Reynolds number and over a vast dynamic range of spatial scales that will depend on stellar mass, chemical composition, and rotation rate.

Despite the great complexity of the underlying physics, \citet[][see also \citealt{Noyes:83}]{Noyes.etal:84} discovered that a remarkably simple pattern emerges from the most elementary  consideration of stellar dynamos. \citet{Noyes.etal:84} noticed that the Ca~II flux vs. rotation period relation for late-type stars exhibits a systematic scatter whose origin depends on stellar spectral type. This scatter could be greatly reduced if, instead of rotation period, the ratio of the convective turnover time to rotation period, $\Prot/\tauc$, is used. In fluid dynamics, this ratio is essentially the Rossby number, $\Ro$, that describes the ratio of inertial to Coriolis forces. With some simple approximations, the Rossby number can be shown to be related to the dynamo number---the ratio of magnetic field generation to diffusion in the convection zone---as roughly $N_\mathrm{D} \propto \Ro^{-2}$ \citep[e.g.][]{Durney.Latour:78,Noyes.etal:84,Mangeney.Praderie:84,Dobson.Radick:89,Montesinos.etal:01,Wright.etal:11}. 

While it has been argued that the Rossby number might not be the fundamental underlying scaling for magnetic activity 
\citep[e.g.,][]{Basri:86,Rutten:87,Stepien:94,Reiners.Mohanty:12,Reiners.etal:14}, 
following the work of \citet{Noyes.etal:84}, many other studies have confirmed and extended the Rossby number-based rotation-activity relation using optical and ultraviolet diagnostics of chromospheric and transition region emission \citep[e.g.,][]{Simon.etal:85,Basri.etal:85,Simon.Fekel:87,Rutten:87,Stepien:94,Cardini.Cassatella:07,Christian.etal:11,Rebassa-Mansergas.etal:13,Houdebine.etal:17,Newton2017,Mittag.etal:18,Pineda2021,Boudreaux2022,Li2024}, and based on X-ray diagnostics of coronal emission \citep[e.g.][]{Mangeney.Praderie:84,Micela.etal:85,Schmitt.etal:85,Maggio.etal:87,Dobson.Radick:89,Jordan.Montesinos:91,Pizzolato.etal:03,Wright.etal:11,Stelzer.etal:16,Gonzalez-Alvarez.etal:19,Pizzocaro.etal:19, Nunez+2022,Magaudda.etal:22,Stassun2024,Shan2024}, including for fully convective M dwarfs  \citep{Wright.Drake:16,Wright.etal:18,Pizzocaro.etal:19,Magaudda.etal:20}. 

Of the ingredients in a Rossby number-based description of stellar activity---rotation period and convective turnover time---only the rotation period is directly observable. The convective turnover time, $\tauc$, is a measure of the timescale of buoyant convective transport and comes from models of stellar interiors. The original work of \citet{Noyes.etal:84} employed the convection zone calculations by \citet{Gilman:80}, where $\tauc$ was evaluated near the bottom of the convective envelope (CE). 

% The rotation rate is a measure of rotational shear, thought to be responsible for converting the stellar poloidal field into a toroidal field, which would form one portion of the dynamo process. Classically, the rotational shear is thought to originate in layers at the base of the convection zone at the tachocline where helioseismology indicates rotational shear is strong \citep{Dikpati.Charbonneau:99,Ossendrijver:03,Weiss.Thompson:09,Charbonneau:10}; however, this is debated, with helioseismology and models indicating the possibility of a near surface shear layer \citep{Brandenburg:2005, Cameron:2023, Vasil:2024}. Convective turbulence is thought to play a role in closing the dynamo loop (although the exact mechanism is unclear \citealt{Charbonneau:2020}) and transport the stellar magnetic field to the surface via buoyancy.

Since the seminal work of \citet{Noyes.etal:84}, several studies have re-examined the convective turnover time, both empirically, by demanding that spectral type dependent scatter in rotation vs.\ activity be minimized \citep[e.g.][]{Stepien:94,Pizzolato.etal:03,Wright.etal:11}, and through numerical stellar evolution models \citep[e.g.][]{Gilliland:85,Rucinski.Vandenberg:86,Rucinski.Vandenberg:90,Kim.Demarque:96,Pizzolato.etal:01,Barnes.Kim:10,Landin.etal:10,Spada.etal:13,Landin.Mendes:17}. More recently, \citet{Corsaro2021} used asteroseismology to calculate convective turnover times for stars in the mass range 0.9--$1.5\,\Msun$ and calibrate a relation as a function of $(B-V)$ and $(G_{\rm BP}-G_{\rm RP})$ colors. Of these later studies, only \citet{Barnes.Kim:10}, \citet{Spada.etal:13}, and \citet{Landin.Mendes:17} probed the lowest mass, fully convective M dwarfs. These stars are of special interest in the study of magnetic activity--rotation relations because of recent doubts that have been cast on the general belief of the dynamo in Sun-like stars originating at the tachocline \citep[e.g.][]{Wright.Drake:16,Wright.etal:18}. 

In this work, we re-calibrate the relation between convective turnover time and mass for low-mass stars using observations of stellar X-ray emission as a function of rotation period, and we use the {\sl Modules for Experiments in Stellar Astrophysics} (\texttt{MESA r11701}) stellar evolution code \citep{Paxton.etal:19} to interpret our results in the context of elementary stellar dynamos. In Section~\ref{s:methodology} we describe the methods we used to estimate convective turnover times empirically and theoretically. In Section~\ref{s:res} we show the results of both methods and compare them. In Section~\ref{s:disc}, we interpret our results for empirical and theoretical convective turnover time, discussing implications for magnetic field generation and the stellar dynamo, as well as caveats. Finally, in Section~\ref{s:conc} we discuss our conclusions.

\section{Methodology}
\label{s:methodology}

Our data and the methodology through which the convective turnover time is derived are described in this section. Our data are drawn from several sources of measured X-ray luminosities and rotation periods. The convective turnover times (and thereby Rossby numbers) are derived both theoretically, i.e., from stellar models, and empirically. The empirical determination follows that of \cite{Stepien:94} and \cite{Wright.etal:11,Wright.etal:18}. The theoretical understanding of the convective turnover time herein, follows the 1D stellar evolution and mixing length theory (MLT) presented in \cite{Henyey:1965, Cox.Giuli:1968}.

\subsection{Empirical Calculation of Convective Turnover Times}
\label{s:tauce}

\subsubsection{Observational Data}
\label{s:obs}

We compiled a catalog of measured ratios of X-ray to bolometric luminosity as well as rotation periods from the literature (see Table~\ref{table:sample_sum} for a summary). This data will (alongside our \texttt{MESA} models are available on Zenodo\footnote{Data \& \texttt{MESA} files: \url{https://zenodo.org/records/13936543}}.
The majority of this sample comprises solar-type stars with masses larger than the fully convective boundary of $\sim 0.35\,\Msun$ retrieved from the work of \citet{Wright.etal:11}. In that study, the X-ray luminosities were computed using the \textit{ROSAT} bandpass ranging from 0.1-2.4 keV, which we also adopt as the reference X-ray luminosity bandpass for the study in hand. 
When necessary, we converted X-ray luminosities of data from other studies to the same energy band using the webPIMMS tool\footnote{\url{https://heasarc.gsfc.nasa.gov/cgi-bin/Tools/w3pimms/w3pimms.pl}}, assuming a plasma/APEC model with $\log_{10} T = 6.8\,$K and $N_\mathrm{H} = 10^{20}$\,cm$^{-2}$ for the conversion. 
%When X-ray uncertainties were not provided by a respective catalog, we assumed a $0.5\%$ uncertainty in logarithmic $\Lx$, corresponding to an $\approx40\%$ uncertainty in linear $\Lx$ (which is consistent with the average uncertainties of measured in the catalogs that provided uncertainties).
If unavailable, we assumed a $20\%$ uncertainty for $\Lx/\Lbol$, consistent with the mean uncertainties reported in other studies \citep[e.g.][]{Stelzer.etal:16, Gonzalez-Alvarez.etal:19, Magaudda.etal:20, Magaudda.etal:22}.

This base sample was extended by new and archival data provided by \citet{Magaudda.etal:20}, who homogenized various different catalogs from the literature \citep{Wright.Drake:16, Wright.etal:18, Stelzer.etal:16, Gonzalez-Alvarez.etal:19} to provide a uniform sample of X-ray activity and rotation properties of M dwarfs. 
We additionally included data for field stars from \citet[][CARMENES/\textit{ROSAT}]{Shan2024}, \citet[][TESS/eROSITA]{Stassun2024}, \citet[][TESS/eROSITA]{Magaudda.etal:22} and \citet[][\textit{Kepler}/\textit{XMM-Newton}]{Pizzocaro.etal:19}, as well as data for the open clusters M34, M35, Praesepe and Hyades from \citet{Gondoin:12}, \citet{Gondoin:13} and \citet[][\textit{ROSAT}/Chandra/Swift observatory/\textit{XMM-Newton}/K2]{Nunez+2022}, respectively.

This resulted in a total of $9344$ sources in the base catalog, out of which, only $1451$ sources remained in our `clean catalog' after applying various quality cuts to these data. 
Below we describe each of the quality cuts, improvements, and considerations we included for our sample.

\textit{Cross-match with \textit{Gaia} DR3:}
We performed a simple cross-match using a $3''$ search radius in the Tool For Operations on Catalogues and Tables \citep[\textsc{Topcat};][]{Topcat}). We found that $98\%$ (i.e. all but $189$ stars) of our sample had a match. This allowed us to update the distances in our sample using the most accurate and precise measurements available, and apply the improvements and quality cuts described below.

\textit{Remove duplicates:} Using the respective \textit{Gaia} ID we identified in total $190$ stars that had more than one X-ray measurement in our sample. From these, we only removed $10$ duplicated sources which had identical measurements of X-ray luminosity, meaning that the same measurement was being included more than once. The duplicated stars with varying values of X-ray luminosity (either due to intrinsic variability of X-ray emission or uncertainty) were kept to improve the calibration of the scatter within the rotation-activity relation. 

\textit{Remove possible binaries:} We subsequently removed stars with Re-normalized Unit Weight Error (\texttt{RUWE}) $>1.4$, which is a parameter provided by the \textit{Gaia} survey, and allows to filter out stars for which the single-star model does not provide a good fit to the astrometric observations\footnote{See \href{https://gea.esac.esa.int/archive/documentation/GDR2/Gaia_archive/chap_datamodel/sec_dm_main_tables/ssec_dm_ruwe.html}{Gaia DR2 documentation}.}. Although extremely useful, the \texttt{RUWE} parameter does not identify tight unresolved binaries. To improve this, we added the cut $\texttt{ipd\_frac\_multi\_peak}>1$, which removes stars that are visually resolved double stars. In addition, we used the flag $\texttt{rv\_amplitude\_robust}$ which indicates the difference between the largest and the smallest radial velocity measured by \textit{Gaia}. This number depends on color and magnitude, becoming larger for fainter stars. Therefore we classified stars as binaries when $\texttt{rv\_amplitude\_robust}>9$ and $G<11$. In total, our binary flag removed $2216$ stars.

\textit{Remove young stars:} As we will discuss in more detail in Section~\ref{ss:mismatch}, the convective turnover time depends among other properties on the stellar age. We therefore removed stars younger than $300$\,Myr for masses $<0.6\,{\rm M_\odot}$, younger than $200$\,Myr for masses in the range $0.6-0.8\,{\rm M_\odot}$, and younger than $100$\,Myr for masses $\geq 0.9\,{\rm M_\odot}$. This cut allows us to keep only stars for which the convective turnover time has stabilized, which does not always agrees with the convergence into the main sequence, as illustrated by Figure~\ref{f:taucm_vsage}. \citet{Stassun2024} included the ages of the stars in their catalog, which we used to apply the cut described above, and for the rest of the sample, we used the Bayesian Analysis for Nearby Young AssociatioNs $\Sigma$ tool \citep[\texttt{BANYAN $\Sigma$},][]{Gagne2018} to estimate the probability that each star belongs to a known moving group from its position, proper motion, parallax and radial velocity from \textit{Gaia} DR3. We removed all the stars that had a probability larger than $0.9$ of belonging to a group which would put them in the ``young" category according to the cut described above. The cut in age removed $6739$ stars, out of which $120$ stars were removed using \texttt{BANYAN $\Sigma$}. We note that we removed most of the stars in the catalog with this age cut. In addition, we used the position in the color-magnitude diagram to identify and remove $106$ stars that belonged to the giant branch by visual inspection.

\textit{Quality cuts:} We applied quality cuts to keep stars with parallax over error $>20$ (equivalent to $5\,\%$ uncertainty) and also uncertainty of $\log_{10}(L_{\rm X}/\Lbol)$ smaller than $0.3$\,dex.

\textit{Metallicity:} Although we do not have metallicity measurements for our sample, we can take into account several factors that justify the sample being mostly made of stars with solar-like metallicity. \citet{Kiman2019} showed that low-metallicity M~dwarfs have a locus below the main sequence in the \textit{Gaia} color-magnitude diagram (CMD). When compared to the CMD of our sample, none of the M~dwarfs in our sample seem to be low-metallicity. In addition, we cross-matched our clean sample with the APOGEE survey \citep{Majewski2017, Smith2021} and found $245$ stars in common ($27\%$). We found that the values for [M/H] are close to solar (between $-0.3$\,dex and $0.3$\,dex) which supports the conclusion we obtained from the CMD position of the stars. 

\textit{Improve mass estimation:} We found that the calculations of stellar masses were inconsistent among the different catalogs. In particular, there are known problems with the mass estimation from \citet[][see \citealt{Jao2022}]{Wright.etal:11}. Therefore, we decided to re-estimate the stellar masses for the complete sample by interpolating within the relations provided by \citet{Pecaut2013}\footnote{\url{https://www.pas.rochester.edu/~emamajek/EEM_dwarf_UBVIJHK_colors_Teff.txt}} to estimate masses from the absolute $G$ magnitude ($M_{\rm G}$). This is a precise mass-luminosity relation that is valid for the widest mass range, in contrast to other calibrations which do not cover a wide range of masses \citep[e.g.,][]{Mann2019,Giovinazzi2022}. %\km{is there a reference for this?}

\begin{figure*}
    \centering
    \includegraphics[width=1.0\textwidth,angle=0]{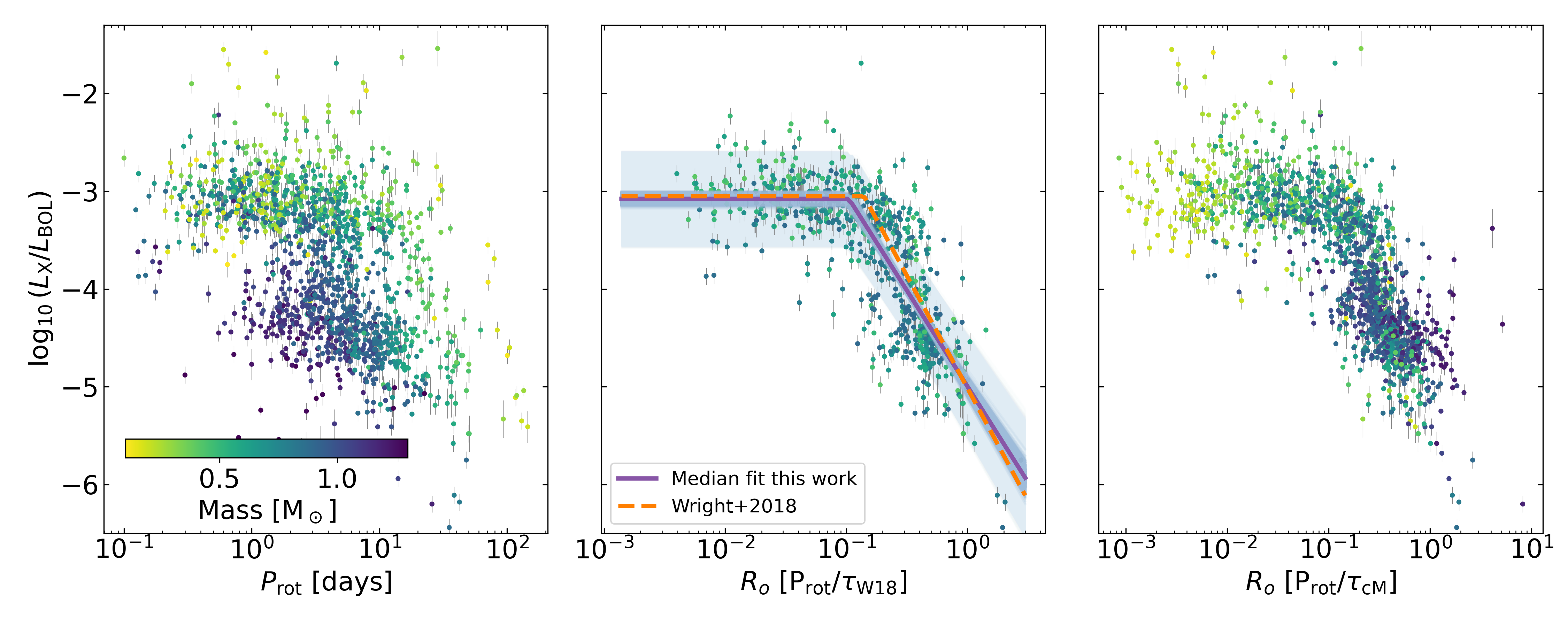}
    \caption{The ratio of the X-ray to bolometric luminosities of our clean sample of stars (see Section~\ref{s:obs} for a description of the sample) as a function of stellar rotation period (left panel), Rossby number calculated using the convective turnover times determined by \citet{Wright.etal:18} for masses between $0.4-0.9\,\Msun$(middle panel), and the theoretical convective turnover time $\tau_{\rm cM}$ at $1$\,Gyr from in this work (see Section~\ref{s:taucm} for the definition; right panel). 
    %In the left panel we show X-ray as a function of stellar rotation period. In the middle panel we show X-ray as a function of Rossby number calculated using the convective turnover times determined by \citet{Wright.etal:18}. And in the right panel we show X-ray as a function of Rossby number calculated using the theoretical convective turnover time $\tau_{\rm cM}$ at $1$\,Gyr from in this work (see Section~\ref{s:taucm} for the definition). 
    The color of each point denotes the stellar mass, as indicated in the color bar. We show our fit to the data in the middle panel: the purple line denotes the fit median using Eq.~\eqref{eq:ro}; $2000$ random samples from the posterior of the parameters are shown in blue shade, and the fit to the scatter of the relation is shown in light-blue shade. The orange dashed line shows the fit obtained by \citet{Wright.etal:18}.}
    \label{f:lxlbol}
\end{figure*}

Our final clean sample comprises $1451$ stars. We show a summary of the compiled sample in Table~\ref{table:sample_sum} and the X-ray luminosities as a function of rotation period for the clean sample of stars color-coded by mass in the left panel of Figure~\ref{f:lxlbol}. 

\begin{deluxetable}{cccc}[ht!]
\tabletypesize{\scriptsize}
\tablecaption{Summary of literature compilation with X-ray and rotation period measurements, before and after quality cuts. \label{table:sample_sum}}
\tablehead{\colhead{Catalog} & \colhead{Stars kept} & \colhead{Total} & \colhead{Reference} \\ 
\colhead{} & \colhead{1451} & \colhead{9344} & \colhead{} 
}\startdata
SK24&629&7545&\citet{Stassun2024}\\ 
W11&386&824&\citet{Wright.etal:11}\\ 
N22&272&440&\citet{Nunez+2022}\\ 
S24&56&113&\citet{Shan2024}\\ 
P19&28&74&\citet{Pizzocaro.etal:19}\\ 
M20&27&54&\citet{Magaudda.etal:20}\\ 
M22&21&223&\citet{Magaudda.etal:22}\\ 
G12&21&41&\citet{Gondoin:12}\\ 
W18&5&16&\citet{Wright.etal:18}\\ 
G13&4&10&\citet{Gondoin:13}\\ 
WD16&2&4&\citet{Wright.Drake:16}\\ 
\enddata
\end{deluxetable}

%\subsection{Convective Turnover Times}
%\label{s:tau}

\subsubsection{Empirical Approach}
\label{s:empiricaltau}

The Rossby number description of stellar activity requires the stellar rotation period in combination with the convective turnover time. The turnover time can be calculated empirically by finding the parameter that reduces the scatter in the activity-rotation relation \citep[e.g.,][]{Noyes.etal:84,Stepien:94,Wright.etal:11,Wright.etal:18}. In this Section, we describe our calculation of the empirical calculation of the turnover time, $\taucE$, following the methods outlined in \citet{Wright.etal:11,Wright.etal:18}. This method allows us to determine $\taucE$ as a function of mass under the assumption that all stars follow the same two-step piece-wise relationship of $\Lx/\Lbol$ as a function of their Rossby number, $\Ro$ (defined as $\Ro=\Prot/\taucE$, where $\Prot$ is the stellar rotation period),
\begin{equation}
\label{eq:ro}
\frac{\Lx}{\Lbol} = 
        \begin{cases}
            C~{\Ro}^{\beta}  & \rm for~Ro > Ro_{\rm sat}\\
            \left(\frac{\Lx}{\Lbol}\right)_{\rm sat} & \rm for~Ro \leq Ro_{\rm sat}\\
        \end{cases}
\end{equation}
where $C$ is a normalization constant, and $\Ro_{\rm sat}$ is the Rossby number at which the transition from saturated to unsaturated regimes occurs. The implication of Eq.~\eqref{eq:ro} is that 
all masses share the same saturated value of $\left(\Lx/\Lbol\right)_{\rm sat}$, and all masses become unsaturated at the same value of $\Ro_{\rm sat}$ and, for higher values of Ro, follow the same power law dependence of $\Lx/\Lbol$ with Ro.%(Ro$^\beta$).

% \citep{Wright.etal:11,Wright.etal:18}. 

In order to determine the values of $\left(\Lx/\Lbol\right)_{\rm sat}$, $\Ro_{\rm sat}$ and $\beta$ for our stellar sample, we fit Eq.~\eqref{eq:ro} to the observations, where $C=\left(\Lx/\Lbol\right)_{\rm sat}/\Ro_{\rm sat}^\beta$.  
To compute Ro, we first used the convective turnover times provided by \citet{Wright.etal:18}. 
As we discuss in Section~\ref{s:tau_comp}, the functional form of the empirical calibration from \citet{Wright.etal:18} differs significantly from the models for high ($>1\,\Msun$) and low ($<0.4\,\Msun$) masses. Therefore, to avoid biasing our fit, we perform this first fit only to masses in the range $0.4-0.9\,\Msun$. The rest of the analysis will be done using the complete sample.
The ratios of $\Lx/\Lbol$ as a function of Ro for the selected mass range of the sample are shown in the middle panel of Figure~\ref{f:lxlbol}. We used the Markov Chain Monte Carlo (MCMC) sampler implemented in \texttt{emcee} \citep{Foreman-Mackey.etal:13} to determine $\beta$, $\left(\Lx/\Lbol\right)_{\rm sat}$, and Ro$_{\rm sat}$. 
In addition, we included a parameter $\sigma$ to fit the scatter of the relation and obtain a more precise estimation of uncertainties. We defined the likelihood of the model as
\begin{equation}
\begin{split}
    &\log\mathcal{L} =  \log(\sigma_{f,i}^2) -0.5\times\\
    &  \sum_i\frac{\left(\log_{10} \left(\frac{\Lx}{\Lbol}\right)_i - \log_{10} \left(\frac{\Lx}{\Lbol}\right)_{i,{\rm model}}\right)^2}{\sigma_{f,i}^2}, \label{eq:likelihood}
\end{split}
\end{equation}
\noindent where $\log_{10} \left(\Lx/\Lbol\right)_{i,{\rm model}}$ is calculated using Eq.~\eqref{eq:ro} and $\sigma_{f}^2$ is the sum of the uncertainty of the X-ray fractional luminosity squared and the characterization of the scatter of the relation squared ($\sigma_{f}^2 = \sigma_{\log_{10} \left(\Lx/\Lbol\right)}^2 + \sigma^2$). To obtain the posterior, we combined the likelihood described above with flat priors for each of the parameters such that $0.01 < {\Ro}_{\rm sat} < 1$, $-5 < \beta < -1$, $-4 < \log_{10}(\Lx/\Lbol)_{\rm sat} < -2$ and $0 < \sigma < 5$.

We report the median of the posterior as the parameter estimate, and the symmetric interval surrounding the median that contains $68.3\%$ of the posterior distribution as the uncertainty. We find $\beta = -1.96 \pm 0.08$, $\Ro_{\rm sat} = 0.11 \pm 0.01$, and $\log_{10} \left(\Lx/\Lbol\right)_{\rm sat} =  -3.08 \pm 0.03$ and $\sigma = 0.40 \pm 0.01$. The parametric fit to the data is also shown in the middle panel of Figure~\ref{f:lxlbol}, where we show the median together with $2000$ random samples from the posteriors of each parameter and the fit to the scatter of the relation. We note that our fit agrees within uncertainties with the results from \citet{Wright.etal:18} shown in an orange dashed line in Figure~\ref{f:lxlbol} ($\beta _{\rm W18} = -2.3 ^{+0.4}_{-0.6}$, $\Ro_{\rm sat, W18} = 0.14 ^{+0.08}_{-0.04}$, and $\log_{10} \left(\Lx/\Lbol\right)_{\rm sat, W18} =  -3.05 ^{+0.05}_{-0.06}$). 

\begin{figure*}
    \centering
    \includegraphics[width=0.8\textwidth]{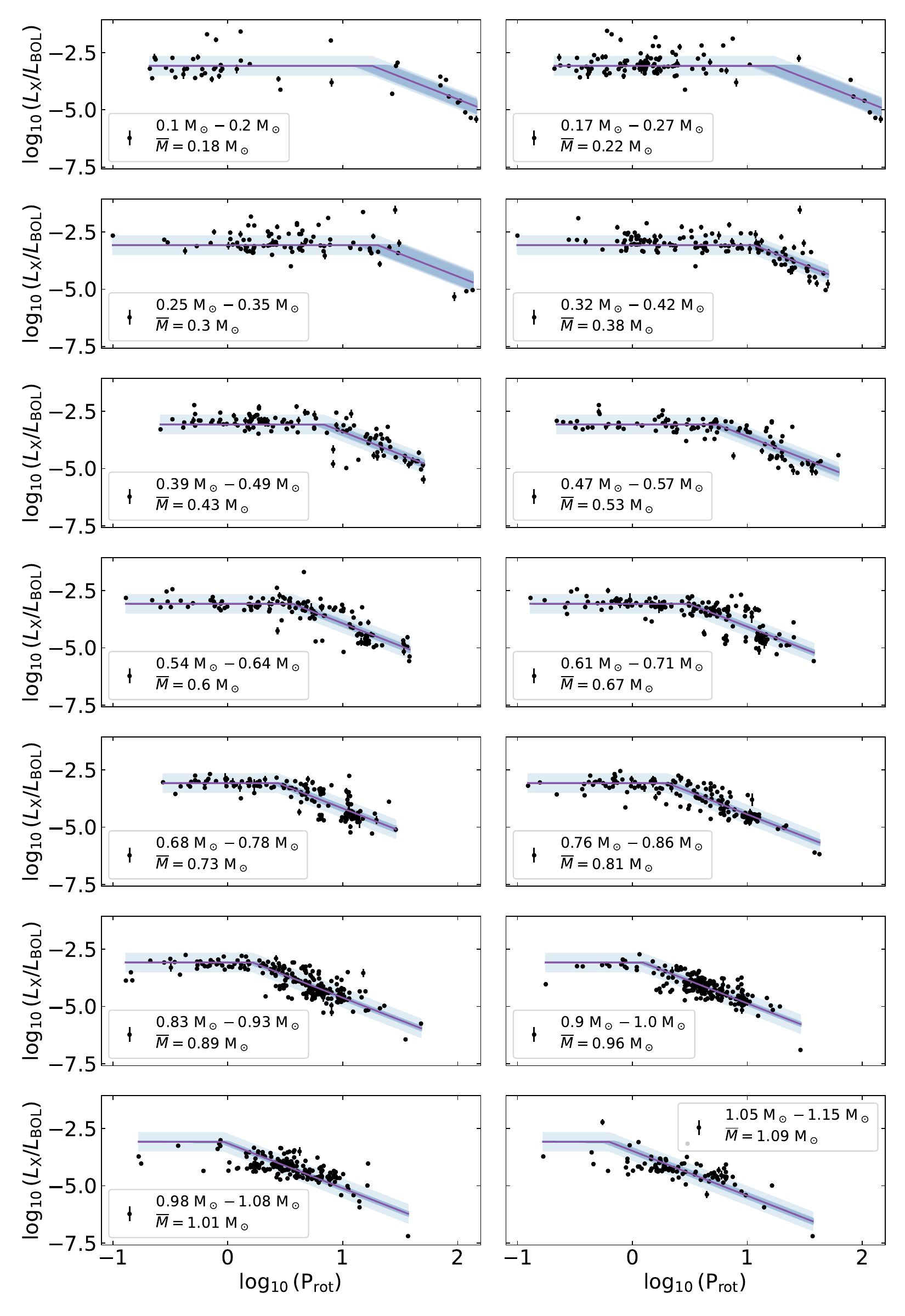}
    \caption{The X-ray luminosity as a function of rotation period (black points) for the different mass bins. The purple solid lines illustrate the best-fit piece wise function (Eq.~\ref{eq:emp_tau}) to locate the rotation period at which stars transition from the saturated regime to the unsaturated regime of X-ray activity. Each panel is labeled with the mass range of the stars included in the fit. We included $2000$ random samples of the posterior for $\Psat$ and a light-blue band that shows the scatter of the relation.}
    \label{f:mass_bin_fit}
\end{figure*}

With $\beta$ and $\left(\Lx/\Lbol\right)_{\rm sat}$ determined, we now fix the parameters $\beta = -1.96$ and $\log_{10} \left(\Lx/\Lbol\right)_{\rm sat} = -3.08$. In addition, we fixed $\sigma=0.40$. These assumptions allow us to determine the rotation period at which saturation occurs for groups of stars of approximately equal mass, independently of the convective turnover time. We divide our sample into $14$ mass bins ranging from $0.1$ to $1.2$~M$_\odot$, with approximately equal numbers of stars in each bin, and fit the equation
\begin{equation}\label{eq:emp_tau}
    {\frac{\Lx}{\Lbol}} = 
        \begin{cases}
            C_p~\Prot^{\beta}  & {\rm for}~\Prot > \Psat\\
            \left(\frac{\Lx}{\Lbol}\right)_{\rm sat} & {\rm for}~\Prot \leq \Psat\\
        \end{cases}
\end{equation}

\noindent using the grid sampling method to obtain the probability distribution of $\Psat$ per mass bin. \texttt{emcee} is significantly slower when we only fit for one parameter, so we decided not to use it in this case. The mass dependant constant $C_p = \left(\Lx/\Lbol\right)_{\rm sat}/\Psat^{\beta}$ can also be defined as $C_p=C/\tau _{\rm cE}^\beta$ using Eq.~\eqref{eq:ro}, which now allows for the determination of $\rm \tau _{\rm cE}$ as a function of stellar mass. We show the individual fits to each mass bin in Figure~\ref{f:mass_bin_fit}. 
By visually inspecting each of the fits, we find that most of them agree with the data. There is a small difference between fit and data for the two highest mass bins: $1.01\,\Msun$ and $1.09\,\Msun$. In these two cases, the slope seems to be less steep in the data. It is possible that the relationship between activity and rotation rate would change towards higher masses, as more massive stars have vanishingly thin convective envelopes (disappearing perhaps after $1.3\,\Msun$. So, it might be expected that the relation should change towards higher masses at some point, as the magnetic dynamo correspondingly becomes ineffective. However we cannot confirm this difference because we do not have enough saturated stars to fully characterize the relation. We note that the difference between data and fit is not large enough to affect the discussions in the rest of the paper.

We confirmed previous results that the rotation period at which saturation occurs clearly increases with decreasing stellar mass \citep[e.g.][and references therein]{Pizzolato.etal:03,Newton.etal:16,Wright.etal:18}. 
As mentioned above, $C_p=C/\tau _{\rm cE}^\beta$, which means that we calculate the convective turnover times using $\tau_{\rm cE}^\beta= C\times \Psat^\beta/\left(\Lx/\Lbol\right)_{\rm sat}$. 
This leaves a constant $C$ to be determined, which will affect the normalization of our calculations of $\taucE$, but not the relative values, as noted also in previous studies \citep{Pizzolato.etal:03,Wright.etal:11}. \citet{Wright.etal:11,Wright.etal:18} decided to choose this normalization, so that their solar value for $\tauc$ agrees with the value in \citet{Noyes:84}, which were theoretically derived. Unfortunately, there is no precise independent estimation of the solar convective turnover time which could be used to normalize our values \citep{Jao2022}, therefore we need to choose between other calibrations. As we are interested in comparing our empirical values with our model values of convective turnover times ($\taucM$, described in Section~\ref{s:taucm}), and with the empirical values obtained by \citet{Wright.etal:18}, we defined $C$ such that $\taucE$ agrees with our theoretical calculations for the mass bin corresponding to $0.96\,{\rm M_\odot}$. This particular mass bin is the closest to the solar-mass bin that has the highest number of stars, which makes the fit to the X-ray versus rotation period more precise (see Figure~\ref{f:mass_bin_fit}). Furthermore, the values for $\tauc$ from \citet{Wright.etal:18} and our theoretical calculations ($14.8$\, and $14.6$\,days, respectively) agree in this bin, which allows the comparison of our empirical $\taucE$ with both calibrations. To calculate the uncertainties of $\taucE$ we did a Monte Carlo propagation of uncertainties of all parameters involved in the calculation ($\Psat$, $\beta$ and $\left(\Lx/\Lbol\right)_{\rm sat}$). We note that although $\beta$ and $\left(\Lx/\Lbol\right)_{\rm sat}$ were fixed to calculate $\Psat$, the uncertainty of the first fit was included in the calculation of $\Psat$ given that we included the $\sigma$ parameter to characterize the scatter of the relation.
The masses and uncertainties for each bin were calculated as the median mass of the bin and the standard deviation, respectively. The resulting convective turnover times calculated with the empirical method are shown in Figure~\ref{f:tau_comp}.

\subsection{Theoretical Calculation of Convective Turnover Times}
\label{s:taucm}

In order to interpret the empirical $\taucE$ values, we calculated theoretical convective turnover times, $\taucM$, using stellar evolutionary models. In making this comparison, we are tacitly accepting the paradigm wherein the empirical turnover time is in fact the same as the theoretical turnover time. In doing so, our aim is to present a physical interpretation of the empirical turnover time by examining correlated stellar properties.

\subsubsection{Stellar Structure Model Calculations}
\label{s:mesamodels}

In order to examine stellar structure model calculations of the convective turnover time, $\taucM$, we used \texttt{MESA}, simulating stars in the mass range $0.1$ to $1.2\,\Msun$ with solar metallicity and a ratio of mixing length to pressure scale height $\rm \alpha_{MLT}=1.82$. Our model profiles as well as full history output at 1, 5, and 14 Gyr, plus inlist and source files are available on Zenodo\footnote{Data \& \texttt{MESA} files: \url{https://zenodo.org/records/13936543}}. Many of our underlying physical assumptions are derived from the \texttt{MIST} models \citep{Choi.etal:16}, based on calibrations described in that text. The value of $\rm \alpha_{MLT}=1.82$ can affect the convection zone size of stellar models, with larger values leading to deeper convection zones. While our choice of $\rm \alpha_{MLT}$ follows from a calibration to solar helioseismic data \citep{Choi.etal:16}, it is similar to the value of 2 chosen by \cite{Noyes:84}. There it was found to minimize the scatter of the activity-rotation relationship and was also noted to reproduce solar values.

These are 1D stellar models evolved to an age of 1\,Gyr, by which time stars in this mass range have settled onto the main sequence. On the main sequence, the convective turnover time has mostly stabilized to a single value and only changes slightly with time. Our models are non-rotating, and while rotation can affect the stellar structure, and thus convective boundaries, we find that such effects are relatively small. Our models are solar metallicity ($Z_{\odot}=0.0142$ from \citealt{Asplund2009}), and while metallicity variations can substantially alter convection zone size, we find the majority of our observed stars are near solar metallicity (see Section~\ref{s:obs}); although, we aim to explore the effect of metallicity variations in greater detail in future work. 

Stars less than approximately $1.3\,\Msun$ possess CEs that start off extremely thin at the higher mass end, and eventually extend all the way to the core at around $M \lesssim 0.35\,\Msun$, as determined via the `Ledoux criterion' \citep{Ledoux:1947} implemented in the \texttt{MESA} models. At $1$~Gyr, stars in this mass range are also primarily on the main sequence, when interior conditions and parameters are mostly changing only very slowly \citep[see, e.g.,][]{Kim.Demarque:96,Landin.etal:10}, and are representative of all but the very lowest mass stars in young open clusters as well as the Galactic disk.

\subsubsection{Theoretical Approach}
\label{s:modeltau}

The convective turnover time is a quantity typically understood through mixing length theory (via \citealt{Henyey:1965}; see \citealt{Joyce.Tayar:23} for a review) that quantifies the timescale of convective motion. It is a local quantity, calculated at a position $r$ within a stellar model as
\begin{equation}
\tauc(r) = \frac{\HP(r)}{\vc(r)}, 
\label{e:tauconv}
\end{equation}
where $\vc(r)$ is the local convective velocity and $\HP(r)$ is the local pressure scale height. Thus, $\tauc(r)$ is the local ratio of the pressure scale height to convective velocity at some location $r$ in the CE, taking on well-defined values in convection zones. It is common to scale $\HP$ by some factor, $\rm \alpha_{MLT}$, the mixing length parameter which typically ranges in value from $1{-}2$, depending on the model. We neglect the scaling factor of $\rm \alpha_{MLT}$, and calculate $\tauc$ as in Eq.~\eqref{e:tauconv}. To calculate the Rossby number ($\Ro=\Prot/\tauc$), one must then decide on the appropriate position $r$ to use when calculating $\tauc(r)$\footnote{$\Prot$ may nominally be considered a function of position as well, but is typically measured at the stellar surface and taken to be as such in modelling}.

For example, as noted in Section~\ref{s:intro}, \citet[][]{Noyes:84} used the convective turnover time calculations of \cite{Gilman:80}. The models of \cite{Gilman:80} (similar to calculations made by \citealt{Durney.Latour:78}) calculated $\tauc$ with $\HP(r)$ evaluated at the bottom of the CE ($r=\rbce$), and $\vc(r)$ at one pressure scale height above $\rbce$ (at $r=\rbce+\HP(\rbce)$; see also \citealt{Gilliland:85}). In the context of stellar dynamos, these choices were made under the consideration that solar and stellar ($\alpha\Omega$) dynamos are thought to reside near the tachocline (near $r=\rbce$). %Furthermore, \cite{Gilman:80}  calculate $\vc(r)$ at $r=\rbce+\HP(\rbce)$, as $\vc(r) \rightarrow 0$ at $r=\rbce$. 

We take Eq.~\eqref{e:tauconv} as the definition of $\tauc$ and test the assumption that the dynamo may lie near $r_{\rm BCE}$ in our analysis. In calculating $\tauc$, there is some subtlety in calculating $\HP$, motivating a slightly different calculation method from \cite{Gilman:80} to calculate $\tauc$ in our work that we describe below. Throughout the text, we use $\taucM$ and $\taucE$ to refer to our model and empirical $\tauc$, respectively.

\subsubsection{Evaluating $\HP(r)$ Above the Convection Zone Base}
\label{s:hp_complications}

Here, we outline complications of calculating $\HP$ in fully convective models and our handling of them. Classically, $\HP$ is calculated as $\HP(r) = \mathrm{P}(r)/g(r)\rho(r)$; $\rm P(r)$ being the local pressure, $\rm g(r)$ the local gravitational acceleration, and $\rm \rho(r)$ the local density, all functions of radius $r$. In fully convective stars $\rbce$ is simply the center of the star. Here $\rm g(0) \rightarrow 0$, causing $\HP(0)$ to diverge under the definition above for fully convective stars. This is problematic for the classical definition of the convective turnover time that evaluates $\HP(r)$ at $r = \rbce$ \citep{Gilman:80,Gilliland:85}. 

In \texttt{MESA}, the divergence of $\HP$ is handled by switching to an alternate definition near the center of the star \citep{Eggleton:71}, $\mathrm{H'_P}(r)=[\mathrm{P}(r)/G\rho^2(r)]^{1/2}$ when $\mathrm{H'_P}(r) < \HP(r)$ (as described in \citealt{Paxton.etal:11}, Section~5.1). 
While this avoids the numerical trap of infinite scale height at the stellar center, it still represents a discontinuity in the treatment of $\HP(r)$ below the fully convective limit. In the fully convective regime, this would then be using the \citet{Eggleton:71} formula for $\mathrm{H'_P}(r)$ and the classical expression for $\HP(r)$ in the partially-convective regime. Additionally, as $\vc(r) \rightarrow 0$ at $r=\rbce$, \cite{Gilman:80} calculated $\vc(r)$ at a different position from $\HP(r)$, i.e., at $r=\rbce+\HP(\rbce)$, one pressure scale height from the bottom of the CE \citep{Gilliland:85}.

%In our models, we use a single position to evaluate both $\HP(r)$ and $\vc(r)$ (called $r_{\HP}$ going forward). Our calculation method of $r_{\HP}$ avoids switching between the \citet{Eggleton:71} and classical form of $\HP$ across the fully convective boundary. %This discontinuity is inherited in the classical calculation of $\vc(r)$ as well, which is evaluated at $r=\rbce+\HP(\rbce)$, one pressure scale height from the bottom of the CE.

%The traditional way of evaluating the location of one scale height above the CE base, $r_{\HP}$, is to add the position of the convection envelope base, $\rm r_{CE,bot}$, and the local scale height at the base, $r_{\HP}=r_{\mathrm{CE,bot}}+\HP(r_{\mathrm{CE,bot}})$ (as in e.g., \citealt{Gilman:80,Noyes:84}).

We choose to evaluate both $\HP(r)$ and $\vc(r)$ -- and thus $\tauc(r)$ -- at the same position, hereafter called $r_{\HP}$. Conceptually, $r_{\HP}$ is similar to where \cite{Gilliland:85} (and \citealt{Gilman:80}) calculated $\vc(r)$ in their models. We utilize
\begin{equation}
r_{\HP}(r) = \rbce+0.5\HP(r)
\label{e:rHP}
\end{equation}
which is the bottom of the CE plus half a (local) pressure scale height. However, rather than setting $r = \rbce$ in Eq.~\eqref{e:rHP} (as in classical works), we solve for the position $r$ to evaluate $\tauc(r)$ as follows.

Starting from the surface of the 1D stellar model, we advance inward cell by cell. At each $i^{th}$ model cell, we evaluate 
\begin{equation}
r_i \lesssim \rbce + 0.5\HP(r_i)
\label{e:rcondition}
\end{equation}
and stop iterating when Eq.~\eqref{e:rcondition} is satisfied, taking $r = r_i$. If we rearrange Eq.~\eqref{e:rcondition}, one may see that we are solving
\begin{equation}
\Delta r_i \lesssim 0.5\HP(r_i)
\label{e:altrcondition}
\end{equation}
where $\Delta r_i \equiv r_i - \rbce$ is an extended distance above the bottom of the convection zone, i.e., the distance from the bottom of the convection zone to the $i^{th}$ cell above it. Eq.~\eqref{e:altrcondition} is satisfied when this distance becomes comparable to half the local pressure scale height in the CE. In our \texttt{MESA} models this locale lies near the bottom of the CE, but always above where $\mathrm{H'_P} < \HP$ and therefore presents a consistent treatment of $\HP$, regardless of convection zone depth. 
We then use $r = r_i$ with Eq.~\eqref{e:tauconv} to estimate our model convective turnover times, $\taucM$.

Going forward, we operate under the framework that the empirical convective turnover time ($\taucE$, described in Section~\ref{s:empiricaltau}) is comparable to the theoretical turnover time. Accordingly, we consider that the empirical value correlates to $\tauc(r)$ evaluated at a particular position ($r$) within the stellar model. The implication is that this position corresponds to the mean location of the magnetic dynamo, as revealed through the activity-Rossby relationship.

%Convective turnover times were extracted as a function of depth from the \texttt{MESA} model computations for the range of stellar masses investigated. We used the convective turnover time definition originally adopted by \citet{Gilman:80} and \citet{Noyes.etal:84}, 

%\textbf{Maybe some mention of the \citet{Hotta.Kusano:21} solar simulation...\\
%\citet{Hotta.Kusano:21} have shown that reproduction the solar differential rotation requires a high-resolution radiative MHD model,and the feedback of strong magnetic fields generated by a small-scale dynamo on thermal convection with the net effect that the convective velocity is suppressed.} 

% divide into Results and discussion. Can eliminate certain speculations
\section{Results}
\label{s:res}

\subsection{Empirical Convective Turnover Times}
\label{s:emp_tau_res}

\begin{figure}[ht]
    \centering
    \includegraphics[width=\linewidth]{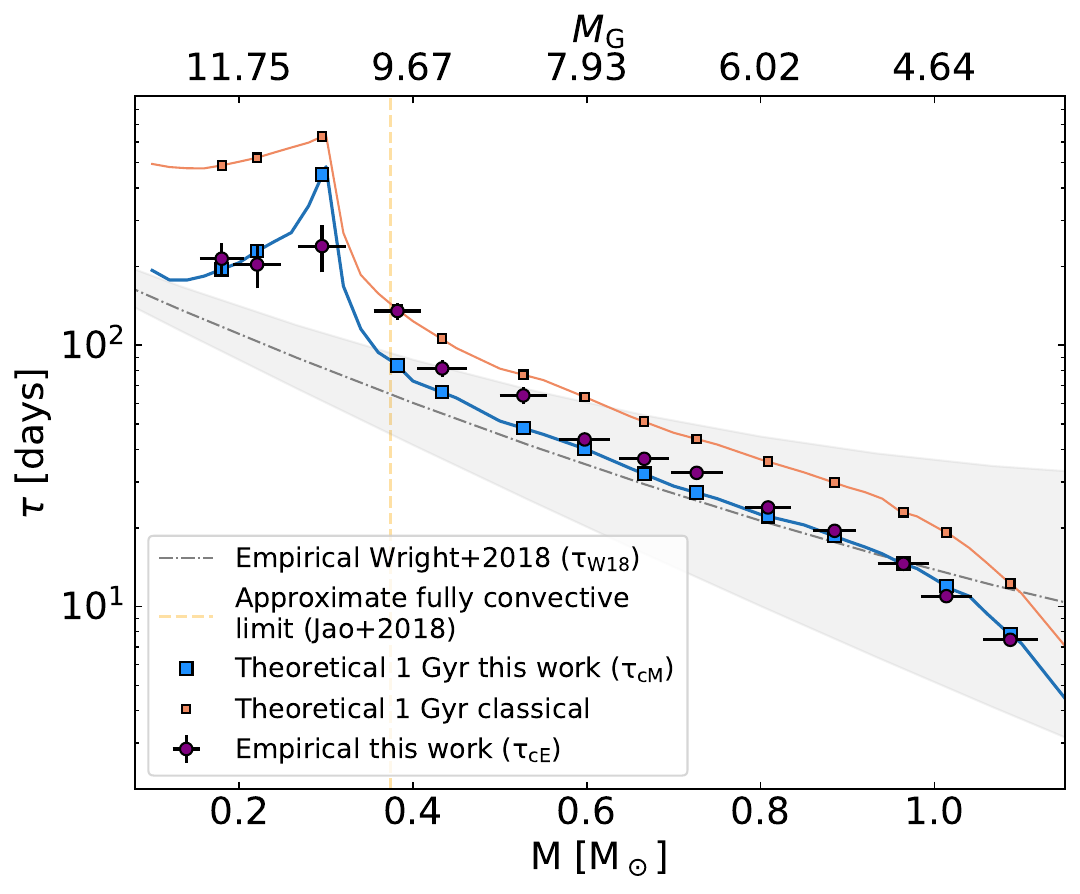}
    \caption{Empirically derived value of $\taucE$ from this work (dark purple points, for each mass bin), compared to the empirically derived $\rm \tau_{\rm W18}$ using the relation of \citet{Wright.etal:18} (gray dashed line and uncertainty in light-gray), and the calculated $\taucM$ from the \texttt{MESA} models for $1$\,Gyr via the new approach described in Section~\ref{s:modeltau} (blue squares) all as a function of stellar mass. In addition, we included the convective turnover times calculated with the classical formalism of e.g., \cite{Gilliland:85} (orange squares).}
    \label{f:tau_comp}
\end{figure}

The empirically-derived $\taucE$ for each mass bin via the methods of Section \ref{s:tauce} are shown in Figure~\ref{f:tau_comp} and Table~\ref{table:results_tau}. We also show a comparison of our empirically-derived $\taucE$ to the empirically-derived $\rm \tau_{\rm W18}$ in \citet{Wright.etal:18}, and to the convective turnover times, $\taucM$, provided by the \texttt{MESA} models described in Section~\ref{s:taucm}. 
The $G$ band absolute magnitude used to estimate masses (Sect.~\ref{s:obs}) is shown on the top axis.

\begin{deluxetable}{cccc}[ht!]
\tabletypesize{\scriptsize}
\tablecaption{Results for our theoretical ($\tau_{\rm cM}$) and empirical ($\tau_{\rm cE}$) calculation of convective turover time. \label{table:results_tau}}
\tablehead{\colhead{Mass [${\rm M_\odot}$]} & \colhead{$\tau_{\rm cM}$ [days]} & \colhead{$\tau_{\rm cE}$ [days]} & \colhead{n}
}\startdata
$0.18\pm0.03$&$196$&$214.59\pm30.90$&$46$\\ 
$0.22\pm0.03$&$229$&$203.91\pm38.73$&$99$\\ 
$0.30\pm0.03$&$451$&$239.77\pm49.66$&$89$\\ 
$0.38\pm0.03$&$84$&$135.36\pm10.04$&$127$\\ 
$0.43\pm0.03$&$66$&$81.54\pm5.96$&$119$\\ 
$0.53\pm0.03$&$48$&$64.29\pm4.71$&$101$\\ 
$0.60\pm0.03$&$40$&$43.61\pm2.50$&$128$\\ 
$0.67\pm0.03$&$32$&$36.77\pm1.79$&$162$\\ 
$0.73\pm0.03$&$27$&$32.49\pm1.64$&$133$\\ 
$0.81\pm0.03$&$22$&$23.95\pm1.26$&$141$\\ 
$0.89\pm0.02$&$19$&$19.49\pm0.73$&$208$\\ 
$0.96\pm0.03$&$15$&$14.58\pm0.53$&$194$\\ 
$1.01\pm0.03$&$12$&$10.96\pm0.41$&$176$\\ 
$1.09\pm0.03$&$8$&$7.46\pm0.42$&$90$\\ 
\enddata
\end{deluxetable}

The main feature in Figure~\ref{f:tau_comp} is the significant increase in convective turnover time around the fully convective boundary ($0.35\,\Msun$) we found with our calculation of $\taucE$, which was not present in the previous empirical calibration $\rm \tau_{\rm W18}$ \citep{Wright.etal:11,Wright.etal:18}. 
In addition, we found that for masses $>0.9\,\Msun$ our calculations of $\taucE$ are smaller than the $\rm \tau_{\rm W18}$, and that this difference increases as the mass increases. We note that, as explained in Section~\ref{s:empiricaltau}, there is a rather arbitrary normalization involved in the calculation of $\taucE$. In this case, the normalization was chosen so $\taucE$ and $\taucM$ agree at $0.96\,\Msun$, which also agrees with $\tau_{\rm W18}$, and allows the comparison of our calculations to both calibrations.

We divided the mass range into $14$ bins with a $0.1\,\Msun$ size. This choice resulted in a smooth trend of convective turnover time as a function of mass, and a similar number of stars in each bin. The final results were insensitive to the exact size and number of bins. 

As described in Section~\ref{s:obs}, we removed binaries from the sample using quality cuts from \textit{Gaia}. However, these cuts are not $100\%$ efficient at removing binaries. Therefore we examined if the results in Figure~\ref{f:tau_comp} changed if we included the binaries, and if we were more strict with the binary cuts (for example, including a cut according to the position in the color-magnitude diagram). We found that the results stay the same and the features did not change significantly. 
To further test our method for systematic errors, we re-ran only the sample from \citet{Wright.etal:18} with masses from the literature to estimate the convective turnover time as a function of mass and were able to reproduced their results. Furthermore, we found that when using the masses estimated using $M_{\rm G}$ with the sample from \citet{Wright.etal:18}, there is a slight trend that shows that the convective turnover time increases at the fully convective boundary, albeit not as clear as in Figure~\ref{f:tau_comp} given that our sample contains a significantly larger number of stars. This shows the importance of estimating accurate masses for this analysis. 

%We fit Eq.~\eqref{eq:ro} to X-ray luminosity as a function of Rossby number where we used the Rossby numbers derived from our empirically derived convective turnover times. We find that  $\rm \beta_{ce} = -1.86 \pm 0.17$, Ro$\rm _{{sat},ce} = 0.14 \pm 0.02$, and  $\rm \log \left(\Lx/\Lbol\right)_{{\rm sat},ce} = -3.15 \pm 0.05$. We find that the Rossby number at which the stars transition from the saturated regime to unsaturated regime and the saturated value of the X-ray to bolometric luminosity are consistent with those we found using the convective turnover times from \texttt{MESA}, while the slope in the unsaturated regime is very marginally steeper.

\subsection{Comparison Between Empirical and Model $\tauc$}
\label{s:tau_comp}

We show our \texttt{MESA} turnover times ($\taucM$) in Figure~\ref{f:tau_comp} as the blue line.
Values of $\taucM$ show generally good agreement with $\taucE$ for (partially convective) masses $0.4\,\Msun<\Mstar<1\,\Msun$, and again for (fully convective) $\Mstar< 0.3\,\Msun$. Qualitatively, the model and empirical values agree fairly well across the entire mass range. Our empirical and model turnover times in particular follow quite closely the relation of \citet{Wright.etal:18} over all masses in terms of slope. However, owing to our larger dataset and different mass estimates, there is a notable difference at lower masses, particularly in the range  $\rm 0.3-0.4\,\Msun$, within which stellar structure is expected to transition from being partially to fully convective \citep[e.g.,][]{Chabrier.Barraffe:1997,Jao2018}. This feature appears as a peak in the empirical convective turnover time values that is matched relatively well (in a qualitative sense) by the theoretical values. 

We show the activity ($\Lx/\Lbol$)-Rossby relation using our theoretical convective turnover time, $\taucM$, in the right panel of Figure~\ref{f:lxlbol}. As expected, the scatter in the relation is significantly reduced compared to $\Lx/\Lbol$ as a function of rotation period (left panel) and of Rossby number of \citet{Wright.etal:18} (middle panel).

We included in Figure~\ref{f:tau_comp} the theoretical calculation of convective turnover times using the classical formalism, described in Section~\ref{s:taucm}. These values are slightly higher than our empirical calculations. However, as noted in Section~\ref{s:empiricaltau}, there is a normalization that was chosen so the empirical values agree with the calibration of our theoretical values $\taucM$ and from \citet{Wright.etal:18}. When normalizing to the classical $\tau$ values, we found good agreement between our $\taucE$ and the classical $\tau$, except for the two least massive bins. Our empirical $\taucE$ agree better with our theoretical $\taucM$ than the classical $\tau$. However, as can be seen in Figure~\ref{f:mass_bin_fit}, these two low-mass bins do not have enough stars to clearly distinguish between the two models. More data is needed for the lowest stellar masses to identify the correct formalism. We note that the differences between our $\taucM$ and classical $\tau$ are small enough that it does not impact the conclusions and discussions in our analysis, with both following a similar trend versus stellar mass.

\begin{figure}
    \centering
    \includegraphics[width=0.47\textwidth]{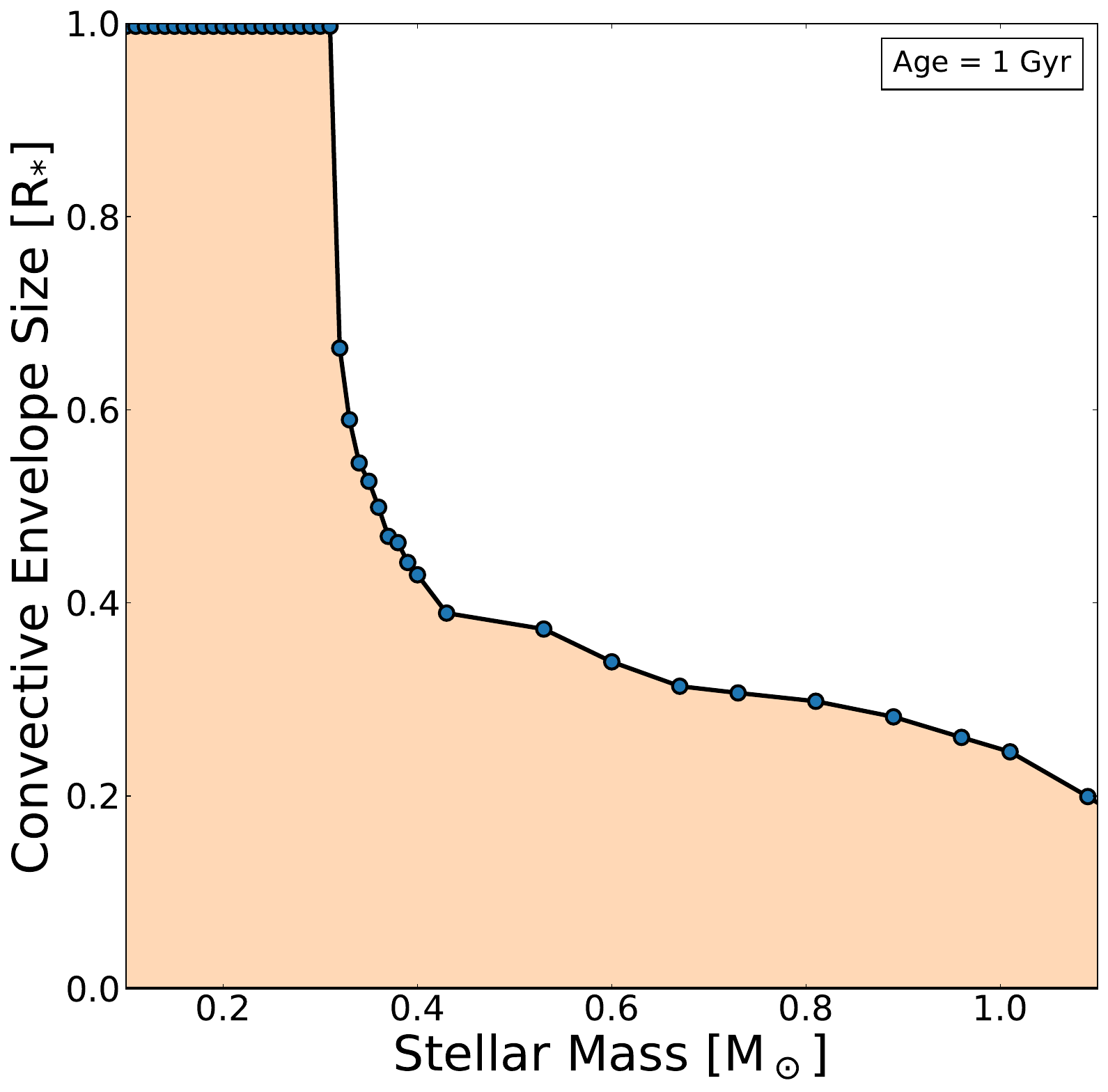}
    \caption{The CE size according to the Ledoux criterion, expressed as a fraction of total stellar radius in late-type stars of solar chemical composition at 1 Gyr as a function of stellar mass. }
    \label{f:convvsmass}
\end{figure}

Our models provide some insight into the nature of the spike in $\taucM$. The CE sizes from our \texttt{MESA} models at 1~Gyr are illustrated in Figure~\ref{f:convvsmass}. Our models suggest that the base of the solar convection zone lies at roughly $0.7\,\Rstar$ at this age, which decreases to about $0.6\,\Rstar$ at $0.4\,\Msun$. At $0.4\,\Msun$, an important transition occurs in which there is a precipitate deepening between $0.4$ to $0.35\,\Msun$ where the convection zone as a function of mass rapidly plunges towards the stellar center. At lower masses still ($M\lesssim 0.35\,\Msun$) our models are fully convective.

As explained in \cite{Chabrier.Barraffe:1997}, the onset of the fully convective limit involves a competition of factors that promote the growth of a convective core and a deepening of the convection zone. When the convective core and outer CE meet, the star may be considered fully convective, leading to the sudden jump in the CE size. This rapid deepening of the CE towards lower masses coincides with the rapid rise in $\taucM$ and $\taucE$.

%At the higher mass end, $\taucE$ implies a larger turnover time than indicated by the model, $\taucM$. 

\subsection{The $0.3\,\Msun$ Peak in $\tauc$ as Seen in Theory}
\label{s:tau_peak}

\begin{figure*}
    \centering
    \includegraphics[width=\textwidth]{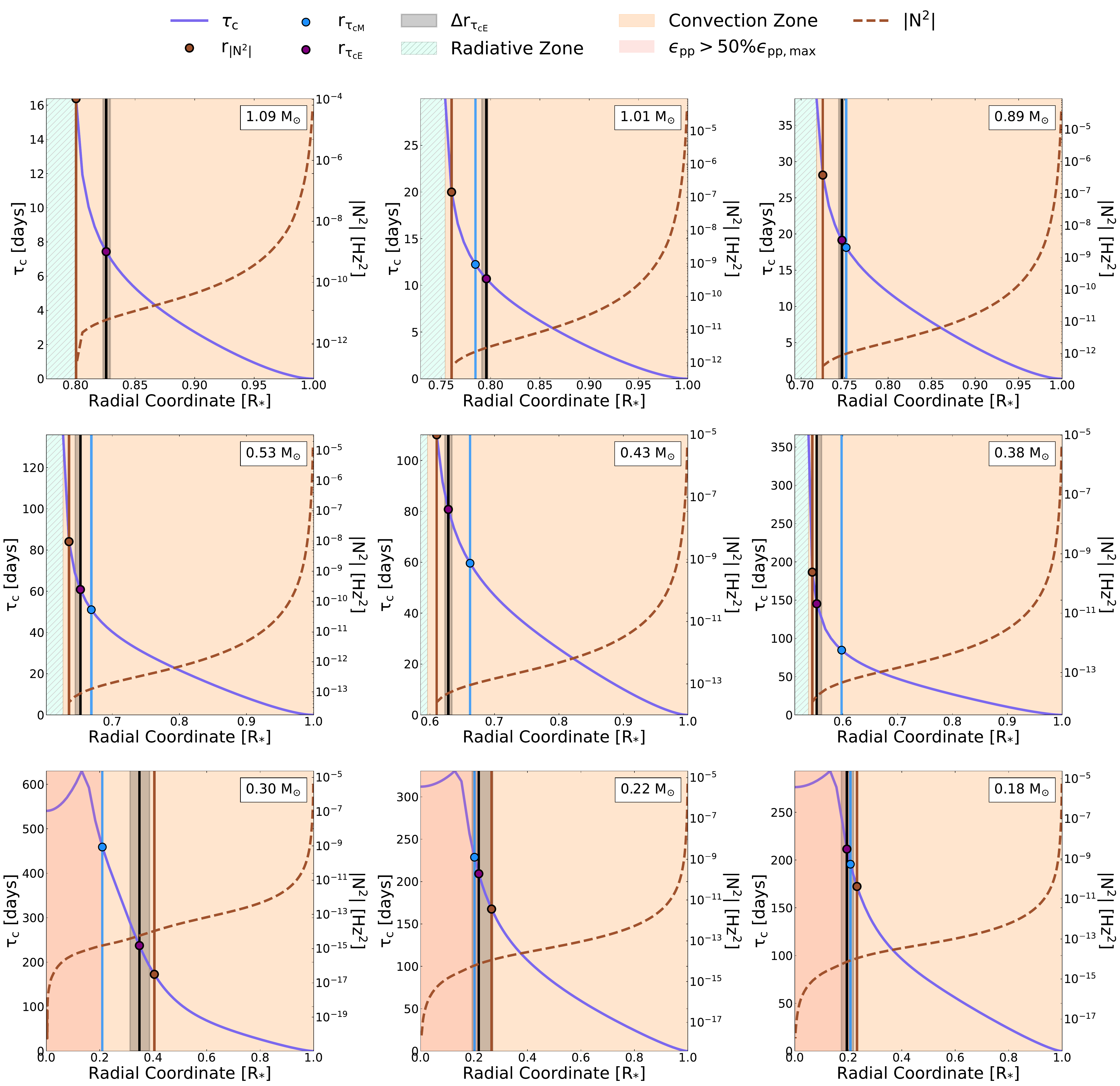}
    \caption{Stellar profiles at an age of 1 Gyr, each panel corresponding to a stellar mass indicated in the upper right, that show the variation of the convective turnover time on the lefthand y-axis (solid blue line) and absolute value of the \bruntvaisala\ frequency (dashed dark orange line) on the righthand y-axis. The stellar surface lies at the righthand side of the plot and the interior lies towards the left. Points along the solid blue curve showing $\tauc(r)$ indicate where values of $\taucM$ (blue dot), $\taucE$ (purple dot) and $\tau_{|N^2|}$ (dark orange dot) exist in the stellar model. Corresponding vertical lines are to show the inferred position in the stellar model. The grey band around the vertical line of $\taucE$ indicates the uncertainty on the inferred location of $\taucE$. Orange regions represent convection zones, and hatched turquoise represents radiative zones. Pink shaded regions indicate where the nuclear energy production rate (via the p-p chain, $\rm \epsilon_{pp}$) is greater than 50\% of its peak value. The inflection points in the $0.18 - 0.3\,\Msun$ model $\tauc(r)$ correspond to where the pressure scale height definition changes to the \citet{Eggleton:71} prescription (see Section~\ref{s:modeltau}). 
    }
    \label{f:convzones}
\end{figure*}

To understand why our calculations of $\taucM$ rise with CE depth, we illustrate the variation with depth of $\tauc(r)$ together with other convection zone properties for different stellar masses in Figure~\ref{f:convzones} (blue line); we also note the location where $\tauc(r) = \taucE$ (the purple dot). This figure will be discussed in further detail in Section~\ref{s:wheretau} below. We note here that, for all masses, $\tauc(r)$ is an increasing monotonic function of depth within the convection zone. 

In the context of Eq.~\eqref{e:tauconv}, $\tauc(r)$ rises with depth due to an increase of $\HP(r)$ with depth, and decrease of $\vc(r)$ towards the bottom of the convection zone. In essence, this convective motion slows with depth due to the weight of overlying material and the effective gravity and buoyancy force being diminished. By design, our calculations of $\taucM$ lie near the bottom of the convection zone (Sec.~\ref{s:modeltau}, blue dot in Figure~\ref{f:convzones}), and so increase as the convection zone deepens towards the fully convective limit.

%We note again that the empirical values $\taucE$ (purple dots in Figure~\ref{f:convzones}), correlate relatively well with our model calculations, $\taucM$. 

As for why $\taucM$ then decreases again for masses $\lesssim 0.3\,\Msun$, we note that once a star becomes fully convective, the bottom of the convection zone becomes fixed (i.e., it is the center of the star). In Figure~\ref{f:convzones} (bottom row), we see that the location corresponding to $\taucE$ begins to settle around $r = 0.2\,\Rstar$. Thus, from our modeling we can say that as this position becomes fixed, and as the stellar radius decreases with stellar mass, the effective CE above this position becomes shallower. As argued above, this causes a decrease in the value of the convective turnover time at this location, leading to the fall in values of $\taucM$ (and presumably $\taucE$) towards lower masses, after the peak, that we see in Figure~\ref{f:tau_comp}. A similar effect occurs at higher masses, approaching $1.3\ \Msun$ as the CE begins to more rapidly shrink in size (as may be seen from Figure~\ref{f:convvsmass})

As previously noted for Figure~\ref{f:tau_comp}, and as may be seen in Figure~\ref{f:convzones} with respect to depth, model-data discrepancies tend to arise near the fully convective limit ($0.3 < \Mstar < 0.4\,\Msun$). 
%At the higher mass end, interpreting $\taucE$ through our models suggests a slightly deeper position than one pressure scale height, closer to the bottom of the convection zone. 
We discuss the nuances related to behavior near the fully convective limit in the following Sections.
We emphasize that the peak in $\taucM$ at $0.3\,\Msun$ (visible in Figure~\ref{f:tau_comp}) is not a unique feature of our \texttt{MESA} models, but has also featured in published convective turnover times by other authors using different codes \citep[e.g.][]{Barnes.Kim:10,Spada.etal:13}. To our knowledge, the origin of this feature has not been discussed before. As discussed further by \cite{Chiti.etal:2024}, such a feature could have strong implications for stellar spin down.

% This can enter into a discussion section
\section{Discussion}
\label{s:disc}

\subsection{Location of $\taucE$ within the Convection Zone}
\label{s:wheretau}

%{\bf JJD: I need to re-arrange this section to get things in a logical flow and order}

The paradigm of the Rossby number approach to interpreting stellar magnetic activity is based on the assumption that dynamo activity is similar across different spectral types. Under this paradigm, the empirical convective turnover time might provide some clues as to the location of dominant dynamo activity in stars with masses, and convection zone properties, quite different to those of the Sun. The similar rotation-X-ray activity behavior in stars on either side of the fully convective limit \citep[e.g., as also in][]{Wright.Drake:16,Wright.etal:18} provides evidence that a tachocline is not a required ingredient for a solar-like dynamo. This raises the question of what convection zone properties drive dynamo activity.

%Here, we make the assumption that $\taucE$ for a given stellar mass correlates with the convective turnover time discussed in dynamo theory, quantifying the contribution of convective dissipation to magnetic field generation in a solar-like dynamo. In this context, $\taucE$ correlates to a location within a theoretical model that minimizes the scatter of the activity-rotation relation, creating the relatively tightly correlated activity-Rossby relationship. Within this paradigm, that location is then assumed to be the location most closely associated with the contribution of convection to the stellar dynamo. 

%Since the traditional definition of the convective turnover time is conceptually problematic for fully convective stars, it is worthwhile examining whether or not there is a more physically motivated reason behind the location of $\taucE$.
%We discuss several convection zone properties identified to at least partly correlate with the location where the empirical turnover time coincides with the turnover time as calculated by our stellar models.

%\subsubsection{Interpreting $r_{\HP}$}

Canonically, the convective turnover time and the location of the dynamo is taken to lie near the bottom of the convection zone (as described in Sec.~\ref{s:modeltau} and~\ref{s:hp_complications}). Our results suggest that the empirical turnover time correlates fairly well with this location, but the pressure scale height itself says little of what the convective properties driving this dynamo action may be. We tested several additional properties in addition to the pressure scale height, that we found roughly track the location of dynamo action implied by $\taucE$, described in the following sections. As a side note, Figure~\ref{f:rcoordvsmass} shows that solar-like stars possess central convective core regions at an age of 1 Gyr. We find that these subside over time, leaving a fully radiative core at the solar age for these models, as expected for solar-like stars.

\subsubsection{The \bruntvaisala\ Frequency and Flux Emergence}\label{ss:brunt}

The \bruntvaisala\, or buoyancy frequency is a quantity describing the frequency of oscillation for a particle displaced in a stable medium. It also serves as an indicator of convective stability in stellar atmospheres. As defined in \texttt{MESA} \citep{Paxton.etal:13}, it may be calculated as 

\begin{equation}
    N^2 = \frac{g^2 \rho}{P}\frac{\chi_T}{\chi_{\rho}}(\nabla_{\rm ad} - \nabla_T + B)
\label{e:bvfreq}
\end{equation}
where $B$ is a quantity accounting for composition gradients, as described in Section~3.3 of that text. The terms $\chi_\rho$ and $\chi_T$ represent the partial derivatives $(\partial \ln P/\partial \ln \rho)_T$ and $(\partial \ln P/\partial \ln T)_{\rho}$, respectively. Hence, the \bruntvaisala\ frequency is directly related to the Ledoux criterion. When $N^2$ takes on positive values, the frequency $N$ is real and leads to oscillatory solutions of motion (stability). However, when $N^2$ is negative, $N$ is imaginary and leads to exponentially growing solutions of motion, i.e., instability. In a stellar model, $N^2$ is positive in radiative zones and negative in convective zones.

We plot $|N^2|$ in the convective zones of our models in Figure~\ref{f:convzones} as the dark orange dashed lines. We find that a position in the stellar model ($r_{|N^2|}$) where $|N^2| > 10^{-14}\ \rm{Hz}^{-2}$ (dark orange points/vertical lines) correlates fairly well with the position inferred by $\taucE$ (purple points/vertical lines). In our higher mass models, $|N^2|$ never gets this low, and this metric is simply placed at the minimum value in the convection zone, near $10^{-13}\ \rm{Hz}^{-2}$ for early M dwarfs and $10^{-12}\ \rm{Hz}^{-2}$ for models representing G-type stars. In Figure~\ref{f:rcoordvsmass}, this location (dark orange squares and line) tracks well with $r_{\rm cE}$.

This may offer some intuition as to what physical processes are at play in the dynamo at $r_{\rm cE}$. This condition suggests that the \bruntvaisala\ growth rate for exponential motion (via unstable buoyant forces in the convection zone) should rise above a threshold before magnetic flux is efficiently transported by convection. Below this threshold the buoyant force would grow at a rate such that it is overwhelmed by the Coriolis force and magnetic flux rises with a trajectory emerging at relatively high latitudes. The values cited above for $|N^2|$ that are found to correlate with $r_{\rm cE}$ are close to those calculated by \cite{DSilva:1995}. For a solar-type star, \cite{DSilva:1995} analytically calculated that $|N^2|$ should be greater than $10^{-12}\ \rm{Hz}^{-2}$ for magnetic flux tubes to rise exponentially in an adiabatic medium and arise at surface latitudes in agreement with sunspots. Those authors derive a condition for threshold values of $|N^2|$ that would scale with mass and could potentially provide a viable alternative approximation for defining where $\tauc(r)$ should be calculated, but is beyond the scope of this work.

The concept of buoyantly rising flux tubes has undergone significant evolution over the last several decades, as reviewed by \cite{Fan:2021}. Initially these models tended to consider magnetic flux being stored and rising from a stable overshoot layer beneath the convection zone and rising to become amplified by rotational shear at the tachocline. As also reviewed by \cite{Charbonneau:2020} and \cite{Cameron:2023}, such interface dynamo models face a number of challenges, and currently produce results in contention with solar observations. Our results, like those of \cite{Wright.etal:18}, suggest that the stellar dynamo may be similar in partially and fully convective stars, the latter in which the tachocline is absent. If the dynamo is situated deep in the CE, as suggested by our findings, then our results lend some evidence to the possibility of a global dynamo operating in the convection zone itself. Our results would also roughly align with findings from \cite{Bice.Toomre:2023} that suggest that a global dynamo may form a reservoir of magnetic flux tubes deep in the convection zone of both partially and fully convective stars, from which these tubes may rise to form active regions on the stellar surface.

Such models are in the vein of those described by e.g., \cite{Nelson:2011,Nelson.etal:2013,Jouve:2013,Nelson:2014,Fan:2014} where shear via turbulent convection may produce magnetic flux tubes, without the need of shear at the tachocline. These simulations tend to find magnetic flux concentrated near the bottom of the CE, which rises to form active regions at the stellar surface. Such simulations are capable of reproducing many observed properties of the solar magnetic field and bypass the difficulties encountered by interface dynamo models. They are however computationally challenging, requiring high spatial resolution to investigate further, as achieved by e.g., \cite{Hotta:2016,Hotta:2020}. Further advancements will be necessary to confirm whether these models can accurately reproduce to solar cycle, and better understand details such as where fields are generated.

\subsubsection{Nuclear Energy Generation Rate }\label{ss:epsnuc}

Both $r_{\rm cM}$ and $r_{|N^2|}$ begin to settle at about 20\% of the stellar radius in fully convective stars, suggesting the dynamo in these stars sits at a fixed fraction of the total stellar radius. As discussed with respect to the \cite{Browning:08} simulations and our models suggesting that convection operates relatively weakly in the centers of fully convective stars, we note further that dynamo action within the stellar core is expected to be relatively weak. Simulations by \cite{Bice.Toomre:2023} also show that the energy available to convection drops below about 0.4 $\Rstar$. Thus, downward advected magnetic fields tend to collect near where this occurs .Motivated by this, we examined a third metric based on the nuclear energy generation rate, in this case due to the proton-proton (p-p) chain reaction (the dominant nuclear reaction in low-mass main sequence stars).  %The motivation for this stems from the reasoning that the dynamo requires energy and so should not be situated below the point in the star where the majority of the energy is generated.

The pink shaded region in Figure~\ref{f:convzones} shows where the specific nuclear energy generation rate, $\rm \epsilon_{pp}$ falls below 50\% of its maximum value. In the outer convection zone of stars with $\Mstar> 0.3\,\Msun$, $\epsilon_{pp}$ is negligible and far from this threshold; however, in fully convective stars with $\Mstar\leq 0.3\,\Msun$, one may see this boundary more clearly as the core comes in to view, with $\rm \epsilon_{pp}$ peaking at the center of the stellar model.

\begin{figure}
    \centering
    \includegraphics[width=0.47\textwidth]{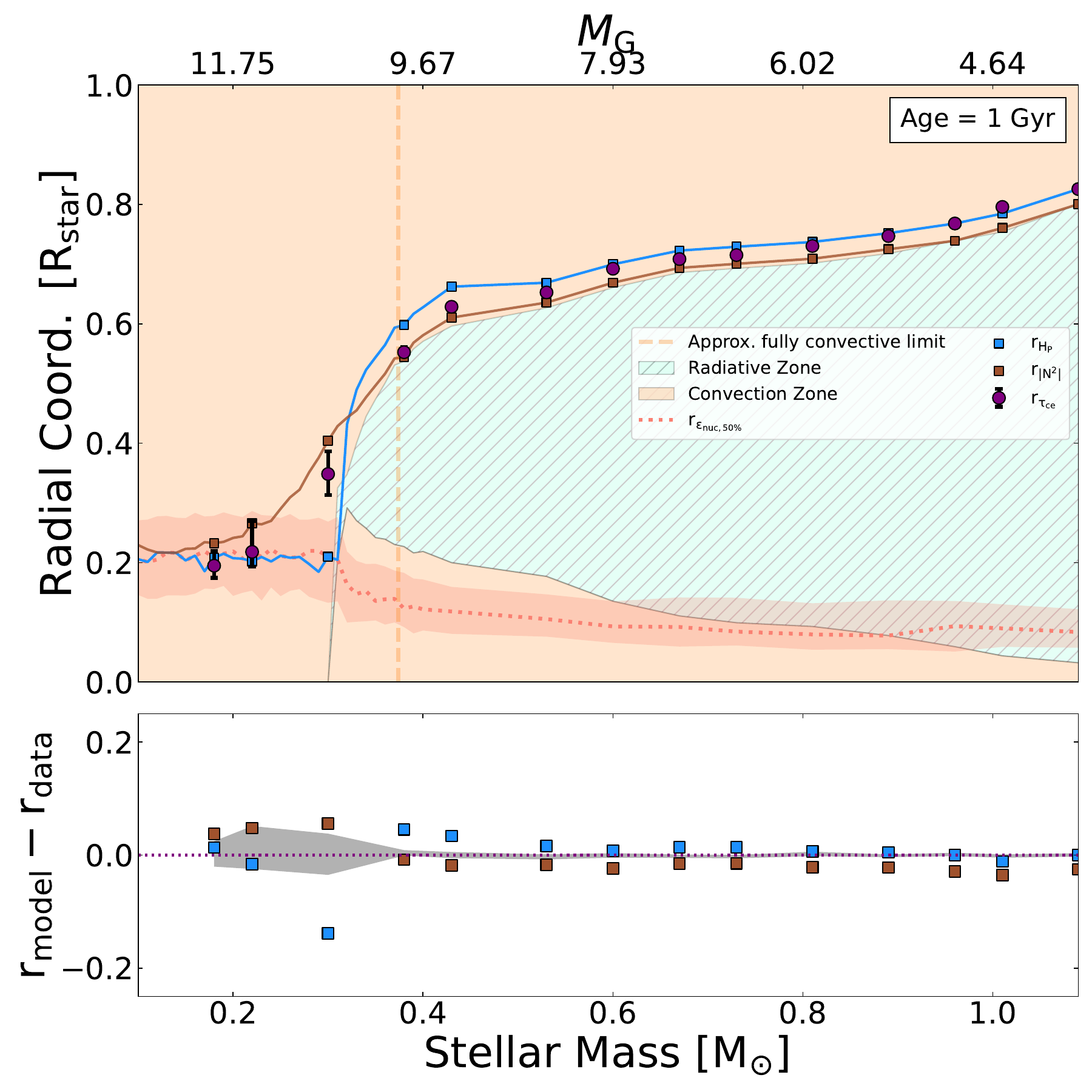}
    \caption{The radial coordinates (where zero represents the stellar center and one the surface) of several physical quantities identified to correlate with $\taucE$ and $\taucE$ itself are plotted for each modeled stellar mass in this study. Error bars correspond to the grey regions of Fig.~\ref{f:convzones}. The bottom panel shows residuals for the radial coordinate between the empirical and model values, with colors corresponding to the legend. See Section~\ref{s:wheretau} for further details.}
    \label{f:rcoordvsmass}
\end{figure}

In Figure~\ref{f:rcoordvsmass}, we plot the metric $\rm r_{\epsilon_{pp},50\%}$, which is the point at which the distribution rises to $30\%$ of its peak value, as the dotted pink line. The shaded region in this case represents the region that spans where the distribution falls from $70\%$ (closer to the center, $\rm r_{\epsilon_{pp},70\%}$) to $50\%$ (closer to the surface, $\rm r_{\epsilon_{pp},50\%}$) of its peak value. We find that the metric $\rm r_{\epsilon_{pp},50\%}$ correlates fairly well with the location of the empirical turnover times in Figure~\ref{f:rcoordvsmass}, roughly agreeing within errors. 

We note that the nuclear energy generation metric is flat in radial coordinate for $M < 0.3\,\Msun$, similar to the pressure scale height metric. Conceivably, a metric based on the bottom of the convection zone and some number of pressure scale heights $\gtrsim 1$, as commonly used, could provide similar results for $M< 0.3\,\Msun$, in agreement with empirical values.
%as long as it sits reasonably outside of the stellar core.

\subsubsection{Relative Convective Luminosity}

Since the $\alpha$-effect in an elementary $\alpha\Omega$ dynamo depends on convection, it might also be expected that dynamo action is not effective where convection is weak. We examined as a function of depth the ratio of convective to total luminosity, $\rm L_{c}/L$, for the range of masses in our study. For masses $\Mstar> 0.3\,\Msun$, the empirical turnover times consistently occur where the ratio of convective to total power is $\rm L_{c}/L \sim 0.35$, confirming that dynamo action does not appear to be associated with zones of weak convection. However,at the lowest masses, $\rm L_{c}/L$ does not reach as low as 0.35 throughout the envelope, rendering any metric for dynamo action based on $\rm L_{c}/L$ problematic to apply.

\subsection{Data-Model Mismatches}\label{ss:mismatch}

Several factors likely contribute to the mismatches seen between our model $\taucM$ and empirical $\taucE$. In particular, these can be seen near the peak of $\tauc$ (at about $0.3\,\Msun$) in Figure~\ref{f:tau_comp}. Here we mention and briefly discuss several possibly contributing factors: the time evolution of the convection zone, the mixing length theory of convection, and data biases.

\begin{figure}[!ht]
    \centering
    \includegraphics[width=0.47\textwidth]{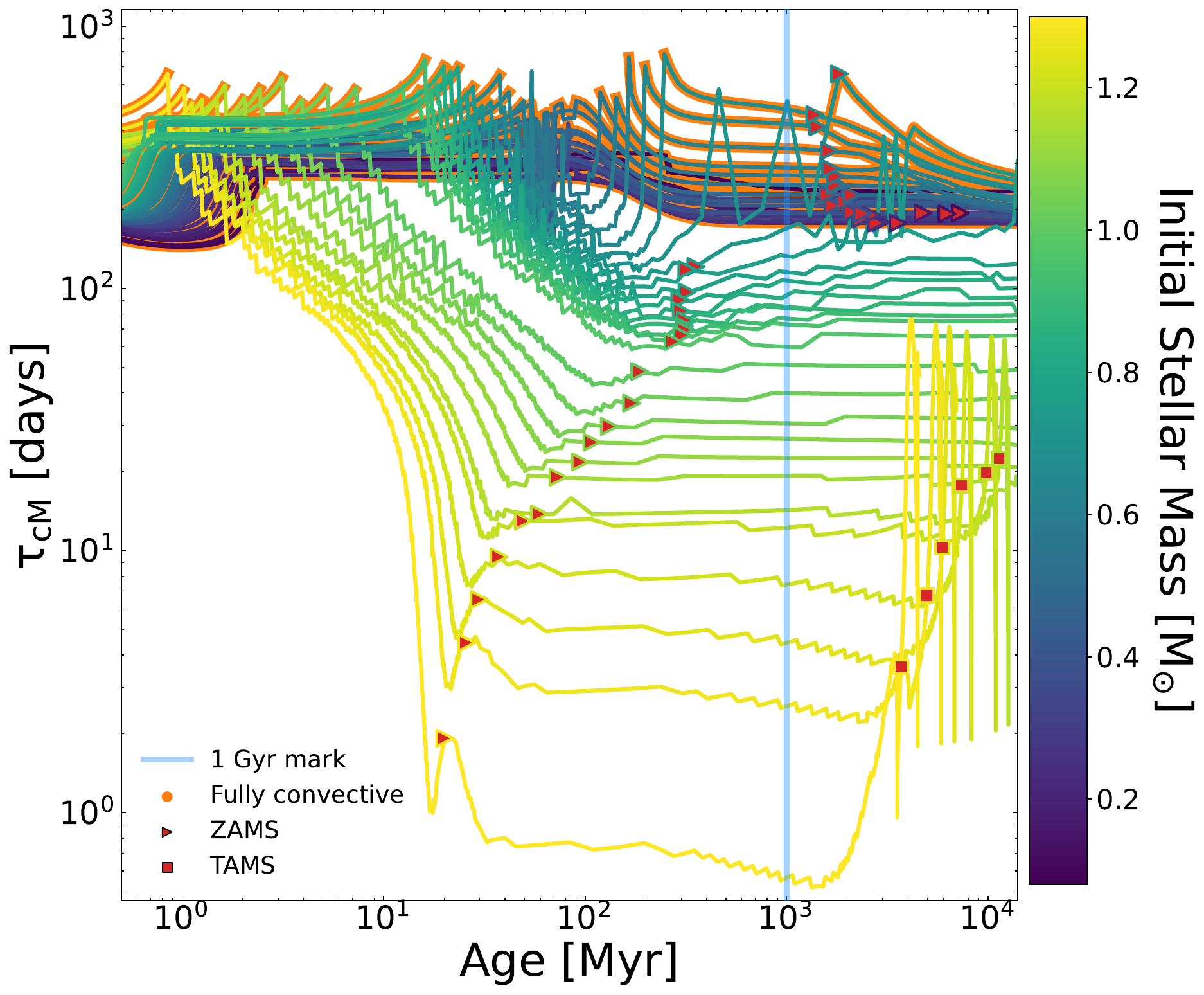}
    \caption{The time evolution of the \texttt{MESA} calculated convective turnover time, $\taucM$ (Section~\ref{s:taucm}). Line colors correspond to stellar masses ranging from $0.08$ to $1.3\,\Msun$. Red triangles indicate the Zero Age Main Sequence (ZAMS) and squares the Terminal Age Main Sequence (TAMS) as defined in \cite{Dotter:2016}. Orange highlights indicate periods of fully convective structure. An age of 1 Gyr is marked with the vertical blue line for reference.}
    \label{f:taucm_vsage}
\end{figure}

Throughout this work we have assumed an age of 1~Gyr for our \texttt{MESA} models, essentially to ensure models (representing field stars, as appropriate for our empirical data) are on the main sequence. In reality, field stars likely exhibit a range of ages (possibly with a majority being $1-8$~Gyr old, e.g., \citealt{Fouesneau.etal:2019, Qiu.etal:2021}). While the convective turnover time is typically not expected to evolve appreciably during the main sequence, there are nonetheless variations that can occur, affecting both stellar structure and consequently derived values of $\rm \tau_{\rm cM}$, as displayed in Figure~\ref{f:taucm_vsage}. 
Here, we constructed a high-resolution grid of \texttt{MESA} models in the span of $0.08$ to $1.3\,\Msun$, evolved for $14$~Gyr to examine the time evolution of $\taucM$ and the stellar structure in more detail. 
It becomes evident that once stars have reached the main sequence, their convective turnover times roughly stabilize before increasing again when reaching the red giant branch.

\begin{figure*}
    \centering
    \includegraphics[width=\textwidth]{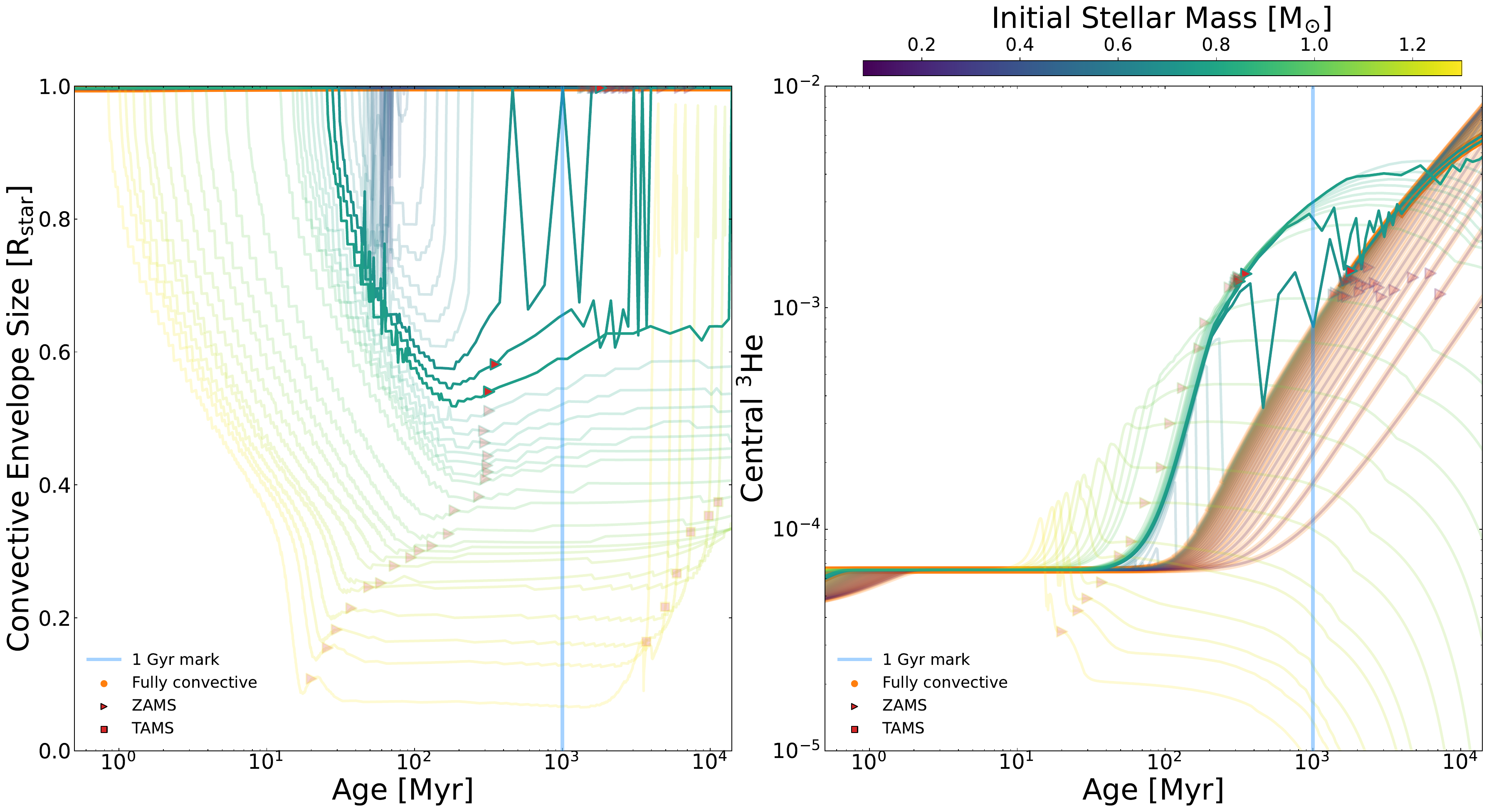}
    \caption{The time evolution of the CE size (CE size, left panel) and central $\rm ^3He$ (right panel) for a range of stellar masses from our \texttt{MESA} models. Plot elements are otherwise as in Figure~\ref{f:taucm_vsage}. Models in the mass range $\rm 0.31-0.33\,\Msun$ exhibiting the convective kissing instability \citep{vanSaders.Pinsonneault:12,Baraffe.Chabrier:2018} leading to relatively dramatic variability in CE size are given greater visibility.}
    \label{f:cesize_he3_vsage}
\end{figure*}

In Figure~\ref{f:cesize_he3_vsage}, we display the CE size (left panel) and central $\rm ^3 He$ (right panel) as functions of time.
We note the periodic fluctuations in convection zone size, which as shown in the right panel of Figure~\ref{f:cesize_he3_vsage}, correspond to a periodic rise and fall of central $\rm ^3He$. Stars in a sub-range of masses within about $\rm 0.3-0.4\,\Msun$ may undergo an instability (dubbed the ``convective kissing instability'' by \citealt{vanSaders.Pinsonneault:12}, but see also \citealt{Baraffe.Chabrier:2018}) that causes the CE of these stars to periodically connect and disconnect from its convective core, due to non-equilibrium $\rm ^3He$ burning. The exact sub-range of stellar masses predicted to exhibit this behavior is model dependant, but in our case it is exhibited by models in the range of roughly $\rm 0.31-0.33\,\Msun$, which are highlighted in Figure~\ref{f:cesize_he3_vsage}. Over time, one may see from Figs.~\ref{f:taucm_vsage} and~\ref{f:cesize_he3_vsage} that the peak feature of $\tauc$ vs. mass will change slightly with time, as the CEs of these stars connect with their convective cores. Further, the peak should become less prominent over time as the thermal structure of these stars relaxes.

As our observed sample of stars likely exhibits a range of ages, rather than the 1 Gyr we compare to in our modeling, we may then expect some discrepancies to arise due to contamination from evolved stars at the high mass end, and phenomena like the convective kissing instability near the fully convective limit. In addition, as can be seen in Figure~\ref{f:mass_bin_fit}, our sample of stars in the mass bin $\rm 0.25-0.35\,\Msun$ (with a median mass of $0.3\,\Msun$), is primarily comprised of stars in the magnetically saturated regime. This mass bin corresponds to the maximum of the peaked feature of Figure~\ref{f:tau_comp}. Thus, with a greater sampling of stars in this mass range, we may see slight shifts in the peak feature of $\taucE$. Furthermore, as discussed by \cite{Chiti.etal:2024}, additional variations on the rotation period measurements of these stars could arise from variable star spot fractions.

Lastly, the theory surrounding the convective turnover time, i.e., mixing length theory, is a 1D approximation of convection, which is in principle a 3D phenomenon in reality. As reviewed by \cite{Joyce.Tayar:23}, MLT may be considered a relatively successful theory, but it ultimately relies on parameters like $\rm \alpha_{MLT}$ that vary study-to-study (typically in the range of $1-2$). With such uncertainty in MLT, precisely predicting where e.g., the onset of full convection occurs, combined with some of the uncertainty discussed above, presents a challenge. This is especially the case considering that the mass range over which full convection takes hold (and presumably where the peak in $\taucE$ occurs) is predicted to be fairly narrow by our models (see Fig~\ref{f:convvsmass}).

\subsection{In the Context of Detailed Dynamo Models}
\label{s:otherdynamos}

Our findings suggest that the stellar dynamo tends to lie deep within the CE of a star. However, recent detailed simulations, such as those of \cite{Vasil:2024}, suggest that solar-like dynamos may viably originate in near surface shear layers (NSSL), perhaps in the outer $5-10\%$ of the star. It is demonstrated that rotational shear, feeding the magneto-rotational instability (MRI) in these outer layers can replicate torsional oscillations \citep{Snodgrass:1985, Voronstov:2002} and subsurface magnetic field amplitudes \citep{Baldner:2009} detected by helioseismology. Thus, the MRI may provide a viable mechanism in the NSSL that drives the global magnetic dynamo.

As reviewed by \cite{Cameron:2023}, the NSSL dynamo, or surface flux transport (SFT) models are founded on various solar observational constraints (see also the review by \citealt{Brandenburg:2005}). As mentioned in those works, a key component of NSSL/SFT models is the process of turbulent pumping within the convection zone. Arguments against NSSL dynamos had suggested that (poloidal) magnetic flux tubes rising to the surface become highly buoyant towards the stellar surface, leading to the possibility that magnetic flux may leave the NSSL before it can be amplified by shearing \citep{Yeates:2023}. Simulations have found that turbulent pumping, where convective downdrafts preferentially transport magnetic flux loops downwards, may counteract this effect, leading to a concentration of magnetic energy near the bottom of the convection zone (see e.g., \citealt{Zhang:2022}).

Bringing the discussion back to the Rossby-activity relation discussed in the present work, the Rossby number, $\Ro = \Prot/\tauc$ quantifies the ratio of magnetic field generation (via shear) to diffusion (via turbulent motion). In this context, our results suggest that the relevant timescale for magnetic diffusion in the dynamo is that of convection near the bottom of the convection zone. As simulations suggest that the bulk of magnetic energy may become concentrated near the bottom of the convection zone as well \citep{Nelson:2011,Nelson.etal:2013,Jouve:2013,Nelson:2014,Fan:2014}, $\tauc$ may then correlate with the timescale on which this energy is brought towards the surface shear layers.

Many studies have approached the problem of dynamos in fully convective stars  \citep[e.g.,][]{Dobler.etal:06,Browning:08,Yadav.etal:2015,Yadav.etal:2016,Brown:2020,Bice.Toomre:20, Bice.Toomre:2023}. Of particular note for the work in hand, \citet{Browning:08} presented 3D MHD simulations of the interior of a fully convective $0.3\,\Msun$ M~dwarf. They found that fully convective stars could generate magnetic fields of kG strength, and that the field was in rough energy equipartition with the convective flows. They also found that the amplitudes and size scales of convective flows varied strongly with radius, with the deep interior convection being comparatively weak. Indeed, the magnetic energy density in their highest resolution model was weak over the inner $30\%$ of the star. Simulations from \cite{Bice.Toomre:2023} suggest fully convective stars may produce magnetic loops through the convection zone, with many collecting in the deep interior before rising to the stellar surface.%, peaking around $\rm 0.6\,\Rstar$---essentially identical to the location of the empirical convective turnover time we find and to which our various metrics correlate in partially convective stars.

Recent studies have found that fully convective stars may often host dipolar magnetic field geometries, as in \cite{See:2020} and as suggested by \cite{Lu:2024} (see also a review by \citealt{kochukhov:21}). The fact that we find dynamo action correlated with the deep stellar interior in this work may corroborate this. As explained in \cite{Browning:08}, simulations of the geodynamo show that slower convection (larger $\tauc$) leads to a stronger influence of rotation on the dynamo. This in turn can lead to magnetic fields generated on larger spatial scales with a larger dipole fraction (see also \citealt{Sreenivasan:2006,Olson:2006,Christensen:2006} for the work on planetary dynamos). Our results suggest that this will tend to be the case for fully convective stars, according to such findings. This is also found in simulations by \cite{Bice.Toomre:2023} of fully convective stars that tend to show greater dipole fractions in their magnetic fields. Ultimately though, this will depend on the rotation rate of the star as well. The nature of fully convective dynamos and the fields they generate is still an active area of research. For instance, simulations by \cite{Brown:2020} found that such stars may produce hemispheric magnetic field topologies, with strong implications for exoplanet habitability and stellar spin down. We look forward to the continuing work in this area being done to improve our understanding of convection and its role in the dynamo under more realistic physical scenarios than the simplified assumptions of MLT allow.

\section{Conclusions}
\label{s:conc}

We have empirically estimated convective turnover times from a compilation of $1451$ stars with measured X-ray/bolometric luminosities from the literature \citep{Wright.etal:11,Wright.etal:18,Wright.Drake:16,Gondoin:12,Gondoin:13,Stelzer.etal:16,Gonzalez-Alvarez.etal:19,Pizzocaro.etal:19,Magaudda.etal:20,Magaudda.etal:22,Nunez+2022,Stassun2024,Shan2024}. Using Bayesian analysis, similar to \cite{Wright.etal:18}, we empirically determined convective turnover times for stars across the mass range $0.1-1.2$\,$\rm \Msun$. Our empirically-derived turnover times show a new feature not found in previous works such as \cite{Wright.etal:18} using the piece-wise $\Lx/\Lbol$-Ro relations of \cite{Wright.etal:11}.

With the data compiled in this work, we find several features in the activity-Rossby relationship that are new compared to \cite{Wright.etal:18}. We find a decrease in $\tauc$ towards $1.3\,\Msun$, associated with decreasing CE size. We also find a sharp rise in the empirical turnover time, seeming to correspond to the stellar mass range where stars become fully convective. We find that theoretical calculations of the convective turnover time, based on mixing length theory, roughly reproduce this feature. Using calculations from \texttt{MESA r11701} stellar models, we examined the theoretical stellar structure corresponding to this feature and why it arises in both the empirical and theoretical values. Our calculations are based on a convective turnover time calculated one half a pressure scale height above the bottom of the convection zone, $\taucM$ (see Section~\ref{s:modeltau}).

Our models suggest that this feature is caused by the sudden deepening of the convection zone as a fully convective structure sets in. This causes a sharp rise in convective turnover times as the bottom of the convection zone plunges towards the stellar core. A subsequent fall in convective turnover time is also seen in both theoretical and empirical values. Our models suggest that this is due to the location of convective motion strongly influencing the dynamo and resultant magnetic activity now being fixed at roughly 20\% $\Rstar$, near the core. Meanwhile the stellar radius shrinks with decreasing stellar mass, effectively allowing buoyancy to become stronger at this position, and $\taucM$ to shrink towards masses lower than about $0.3\,\Msun$.

Our results lend support to the possibility that the magnetic dynamos of stars with $0.1 < \Mstar \leq 1.2\,\Msun$ may operate similarly. At least, here it is suggested that dynamo action driving magnetic activity measured by $\Lx/\Lbol$ associated with convection is seated deep within the outer convection zone, even when the star becomes fully convective (in which case it is near, but above, the stellar core). Our results are consistent with simulations suggesting that the bulk of magnetic energy may become concentrated near the bottom of the convection zone \citep{Brandenburg:2005,Nelson:2014,Fan:2014,Zhang:2022,Bice.Toomre:2023,Cameron:2023}. This further corroborates evidence found by \cite{Wright.etal:18} that suggests partially and fully convective stellar dynamos have similar mechanics. Likewise, this corroborates results from simulations \citep[e.g.,][]{Munoz.etal:2009,Brown.etal:2010,Nelson.etal:2013,Fan:2014,Yadav.etal:2016,Hotta:2020} that fully convective stars may be capable of operating a dynamo similar to partially convective stars, operating throughout the convection zone. 

%\textcolor{red}{We additionally compared Bayesian information criteria for an increasing $\Lx/\Lbol$ ratio in the saturation regime, from \cite{Reiners:14}, finding that it is statistically favored compared to the relation from \cite{Wright.etal:11}, which also supports findings by \cite{Magaudda.etal:20}. Comparing the model of \cite{Blackman.Thomas:15} to the piece-wise model of \cite{Wright.etal:11}, we find that the latter is statistically favored as well.}

% TO ADD?
% More on e.g., the likelihood and priors?
% Mention caveats regarding that these models are non-rotating? Metallicity? Maybe a small statement that we will explore this in the future.

\acknowledgments

We extend warm thanks to Jan Eldridge for insights and discussion regarding the Eggleton pressure scale height formula.
SG acknowledges funding from Northwestern University through a CIERA Postdoctoral Fellowship.
RK was supported in part by the Simons Foundation (668346, JPG). KM was supported by NASA {\it Chandra} grants GO8-19015X, TM9-20001X, GO7-18017X, and HST-GO-15326. 
AAM was supported by NSF Graduate Research
Fellowship, Grant No.~DGE1745303.
We would also like to thank Bill Paxton and the \texttt{MESA} community for making the stellar evolution code on which this work is based possible. 
This work has made use of data from the European Space Agency (ESA) mission
{\it Gaia} (\url{https://www.cosmos.esa.int/gaia}), processed by the {\it Gaia}
Data Processing and Analysis Consortium (DPAC,
\url{https://www.cosmos.esa.int/web/gaia/dpac/consortium}). Funding for the DPAC
has been provided by national institutions, in particular the institutions
participating in the {\it Gaia} Multilateral Agreement.


%\vspace{5mm}
\facilities{This research was supported in part through the computational resources and staff contributions provided for the Quest high performance computing facility at Northwestern University which is jointly supported by the Office of the Provost, the Office for Research, and Northwestern University Information Technology.}


\software{\texttt{MESA r11701} \citep{Paxton.etal:19};  \texttt{scipy} \citep{SciPy-NMeth2020}; \texttt{numpy} \citep{harris2020array}; \texttt{matplotlib} \citep{Hunter2007}; \texttt{astropy} \citep{astropy2013,astropy2018,astropy2022}; \texttt{emcee} \citep{Foreman-Mackey:2013}}



%\clearpage




\clearpage

\end{document}